\documentclass[aps,showpacs,twocolumn,twoside,pra,10pt]{revtex4-1}
\usepackage{amsmath} 
\usepackage{amssymb}
\usepackage{graphicx}
\usepackage{color}      
\usepackage{hyperref}   
\def\ket#1{|#1\rangle}

\def\scal#1#2{\langle#1|#2\rangle}
\def\matr#1#2#3{\langle#1|#2|#3\rangle}
\def\abs#1{\left\lvert#1\right\rvert}

\def\ave#1{\langle#1\rangle}

\def\dis#1{\langle\!\langle#1^2\rangle\!\rangle}
\def\={\!=\!}
\def\>{\!>\!}
\def\<{\!<\!}
\def\-{\!-\!}
\def\+{\!+\!}

\def\abs#1{\left|#1\right|}

\def\uvo#1{\lq\lq #1\rq\rq}
\def\Eqref#1{Eq.\,\eqref{#1}}

\begin{document}

\title{Quantum quench dynamics in Dicke superradiance models}

\author{Michal Kloc}
\email[E-mail address: ]{kloc@ipnp.troja.mff.cuni.cz}
\author{Pavel Str\'{a}nsk\'{y}}
\author{Pavel Cejnar}

\affiliation{Institute of Particle and Nuclear Physics, Faculty of Mathematics and Physics, Charles University, V\,Hole{\v s}ovi{\v c}k{\'a}ch 2, Prague, 18000, Czech Republic}

\date{\today}

\begin{abstract}
We study the quantum quench dynamics in an extended version of the Dicke model where an additional parameter allows a smooth transition  to the integrable Tavis-Cummings regime.
We focus on the influence of various quantum phases and excited-state quantum phase transitions (ESQPTs) on the survival probability of the initial state.
We show that, depending on the quench protocol, an ESQPT can either stabilize the initial state or, on the contrary, speed up its decay to the equilibrated regime.
Quantum chaos smears out the manifestations of ESQPTs  in quench dynamics, therefore  significant effects can only be observed in integrable or weakly chaotic settings.
Similar features are present also in the post-quench dynamics of some observables.

\end{abstract}

\maketitle

\section{Introduction}

The advent of quantum simulators opened routes to laboratory realizations of artificial quantum systems designed either to implement certain utilizable functions, or to demonstrate some fundamental principles \cite{Gar16,Geo14}.
Examples include efforts to built an efficient quantum computer (see, e.g., Ref.\,\cite{Ghe17}) or experiments with quantum phase transitions (see, e.g., Refs.\,\cite{Gre02,Bau11,Kli15}).
An important mode of use of quantum simulators lies in probing the dynamics of quantum systems far from thermodynamical equilibrium \cite{Pol11,Eis15}.
This is often achieved via a protocol called quantum quench, which in its sudden form consists in the following steps \cite{Sen04,Cal06,Sil08}: 
(a) take a system described by Hamiltonian $\hat{H}_{\rm i}$  and prepare it in an eigenstate $\ket{\psi_{\rm i}}$, 
(b) suddenly switch the Hamiltonian to $\hat{H}_{\rm f}$, and 
(c) with increasing time $t$, measure the probability $P(t)$ of finding the initial state $\ket{\psi_{\rm i}}$ in the state evolved from it by $\hat{H}_{\rm f}$.
It is usually assumed that the initial and final Hamiltonians are members of the same family depending on an external parameter $\lambda$, so $\hat{H}_{\rm i}\equiv \hat{H}(\lambda_{\rm i})$ and $\hat{H}_{\rm f}\equiv\hat{H}(\lambda_{\rm f})$ with $\lambda_{\rm i}$ and $\lambda_{\rm f}$ denoting initial and final parameter values.

The evolution of the initial-state survival probability $P(t)$ is entirely encoded in the energy distribution of the initial state expressed in the final Hamiltonian eigenstates.
At first, $P(t)$ drops with a rate related just to the final energy dispersion.
However, at later stages of the evolution, rather complex dynamical regimes occur which gradually disclose more and more subtle details of the final energy distribution.
Correlations between individual final energy levels and the corresponding occupation probabilities become apparent at these stages \cite{Bor16}.
Although the medium- and long-time evolution is usually characterized by a very low average of the survival probability, sharp peaks of $P(t)$ are repeatedly observed, indicating sudden revivals of the initial state.

The quench dynamics in complex quantum systems currently represents a subject of intense theoretical and experimental research, see, e.g., Refs.\,
\cite{Gre02,Kli15,Eis15,Sen04,Cal06,Sil08,Bor16,Cra08,Gra10,Fer11,Mon12,Cam14,Tav16,San16,Hey17,Jaf17,Ber17,Tav17,Tor18,Mit18}.
An important direction of this research is aimed at dynamical imprints of various forms of quantum criticality that emerge in the infinite-size limit of some systems.
For example, the so-called Dynamical Quantum Phase Transition (DQPT) shows up as a non-analyticity observed directly in the survival probability as a function of time \cite{Kli15,Hey17}. 
Also relevant for the quench dynamics are the concepts of ground-state Quantum Phase Transition (QPT) \cite{Sac11,Car11} and Excited-State Quantum Phase Transition (ESQPT) \cite{Cej06,Cap08,Str08,Str15,Str16}.
Since these are rooted in non-analytic properties of the system's stationary states (ground or excited states) taken as a function of the control parameter, their effect in quench dynamics is not seen as a sharp-time anomaly like in the DQPT case, but rather shows as a qualitative change of the quench dynamics with varying difference  $\Delta\lambda=\lambda_{\rm f}-\lambda_{\rm i}$.
The changes appear if the parameter shift connects different quantum phases of the system.
In spite of numerous theoretical efforts to clarify the relations between (ES)QPTs and quench dynamics (see, e.g., Refs.\,\cite{Sen04,Cal06,Sil08,Cra08,Gra10,Mon12,Cam14,Jaf17,Mit18} for QPTs and \cite{Fer11,San16,Ber17} for ESQPTs), the problem still remains open for further investigation.

In this paper, we address this problem, particularly that of the ESQPT influence, by analyzing the quench dynamics in a model generalizing the Dicke superradiance phenomenon in cavity QED systems \cite{Dic54,Hep73,Wan73}.
The model contains a QPT \cite{Ema03} and several types of ESQPTs \cite{Fer11,Bra13,Bas14,Bas16a,Lob16,Klo17a,Klo17b}.
Its ground-state critical behavior was demonstrated experimentally \cite{Bau11}.
We show that the effect of  ESQPTs on the quench dynamics strongly depends on the ESQPT type (in the sense of the classification described in Ref.\,\cite{Str16}) and on the quench protocol (selection of the initial state and size of the parameter change).
The effect is most pronounced in the integrable Tavis-Cummings regime \cite{Tav68}, in which the dynamics becomes effectively determined by a single degree of freedom, but can be observed also in the full (non-integrable) regime.
 
The plan of the paper is as follows:
In Section \ref{TB}, we introduce an extended version of Dicke model (Sec.\,\ref{EDM}), that will be employed in this work, and outline the general theoretical background on the quench dynamics (Sec.\,\ref{QQD}).
In Section \ref{NUR}, we describe numerical results on the quench dynamics gained within the model in its integrable (Sec.\,\ref{F1}) and non-integrable regimes (Sec.\,\ref{F2}).
We focus on two types of quench protocols named forward and backward protocols.
Section\,\ref{SUM} brings a brief summary.

\section{Theoretical background}
\label{TB}
 
\subsection{The extended Dicke model}
\label{EDM}

The Dicke model \cite{Dic54} was formulated to describe an idealized interaction of an ensemble of $N$ two-level atoms with one-mode electromagnetic field.
The original intention was to demonstrate the dynamical superradiance phenomenon, i.e., coherent radiation induced by collective behavior of atoms in the regime of weak atom-field coupling.
Later it turned out that in the strong coupling regime, the model exhibits another form of superradiance, namely a thermal phase transition \cite{Hep73,Wan73} and, in the zero-temperature limit, the corresponding QPT \cite{Ema03} to an equilibrium superradiant phase.  
A route to laboratory realizations of the cavity QED system with strong atom-field coupling was proposed in Ref.\,\cite{Dim07} and led to successful experiments with the superradiant QPT reported in Ref.\,\cite{Bau11}.

We use here the so-called Extended Dicke Model \cite{Bra13,Bas14,Klo17a,Zhi17} with the Hamiltonian (we set $\hbar=1$)
\begin{equation}
\hat{H}=\omega\ \hat{b}^{\dag}\hat{b}+\omega_0\hat{J}_z+\frac{\lambda}{\sqrt{N}}\left[\hat{b}^{\dag}\hat{J}_-+\hat{b}\hat{J}_++\delta\left(\hat{b}^{\dag}\hat{J}_++\hat{b}\hat{J}_-\right)\right]
\label{H}
\,.
\end{equation}
Here, $\hat{b}^\dagger$ and $\hat{b}$ stand for creation and annihilation operators of the bosonic field (photons) while $\hat{J}_{\pm}$ and $\hat{J}_z$ represent collective quasi-spin operators affecting the ensemble of atoms.
These are written as $\hat{J}_\bullet=\sum_{k=1}^{N}\hat{\sigma}^k_\bullet/2$ (with the symbol $\bullet$ standing for $z,+,-$), where $\hat{\sigma}^k_\bullet$ is the Pauli matrix acting on the $k$th atom whose lower and upper states read as $\left(\smallmatrix 1\\ 0\endsmallmatrix\right)$ and $\left(\smallmatrix 0\\ 1\endsmallmatrix\right)$, respectively [so, for example, $\hat{\sigma}^k_+/2=(\hat{\sigma}^k_x+i\hat{\sigma}^k_y)/2$ excites the $k$th atom from the lower to the upper state etc.].
The first two terms in Eq.\,(\ref{H}) represent the free field (with a single boson energy $\omega$) and the free atoms (with a single atom excitation energy $\omega_0$), while the third term describes an atom-field interaction with an overall strength $\lambda\in[0,\infty)$ and an additional parameter $\delta\in[0,1]$ which scales the so-called counter-rotating terms.
The variation of $\delta$ induces a smooth crossover from the Tavis-Cummings regime \cite{Tav68} with $\delta=0$ (full neglect of the counter-rotating terms) to the original Dicke regime \cite{Dic54} with $\delta=1$.

Note that the use of collective quasi-spin operators in the interaction term is based on the assumption that the size of the atomic ensemble is much smaller than the wavelength of radiation, so that the interaction is uniform for all atoms.
Since $\hat{J}^2$ commutes with the Hamiltonian (\ref{H}), the dynamics does not mix subspaces with different quantum numbers $j$ of $\hat{J}^2$.
We therefore select a single-$j$ subspace, namely that with the maximal value $j=N/2$, which is fully symmetric with respect to the exchange of atoms (subspaces with lower values of $j$ appear in numerous replicas differing by the type of exchange symmetry) \cite{Cej16}.
So all physical states $\ket{\psi}$ can be expressed as linear combinations of the basis states $\ket{n}\ket{m}$, where $n=0,1,2,\dots$ is the number of bosons and $m=-j,-j+1,\dots,+j$ the quasi-spin $z$-projection. 

The classical limit of Hamiltonian (\ref{H}) can be realized in terms of two pairs of canonically conjugate coordinate and momentum variables, hence the model has in general  two degrees of freedom: $f=2$ \cite{Ema03,Bas14}.
One of them is associated with the field states; the corresponding part of the phase space is a plane.
The other represents the collective atomic states; their phase space is the Bloch sphere.
Details can be found, e.g., in Refs.\,\cite{Klo17a,Klo17b}.
The minimum of the classical energy landscape in the whole phase space, i.e., the lowest stationary point of the system, corresponds to the energy of the quantum ground state in the $N\to\infty$ limit.
It can be written as
\begin{equation}
\frac{E_{\rm g.s.}}{\omega_0j}=\left\{\begin{array}{ll}
-1 & {\rm for\ }\lambda\in[0,\lambda_{\rm c})\,,\\
-\frac{1}{2}\left(\frac{\lambda_{\rm c}^2}{\lambda^2} +\frac{\lambda^2}{\lambda_{\rm c}^2}\right) & {\rm for\ }\lambda\in[\lambda_{\rm c},\infty)\,,
\end{array}
\right.
\label{Egs}
\end{equation}
where the critical parameter value
\begin{equation}
\lambda_{\rm c}=\frac{\sqrt{\omega \omega_0}}{1+\delta}
\label{LamC}
\end{equation}
sets a second-order ground-state QPT, where $d^2E_{\rm g.s.}/d\lambda^2$ changes discontinuously.

Higher (unstable or quasi-stable) stationary points of the classical Hamiltonian demarcate the ESQPTs of the model.
Detailed analyses can be found in Refs.\,\cite{Bas14,Klo17a}. 
One of the ESQPT critical borderlines in the plane parameter $\lambda$ $\times$ energy $E$ can be written as
\begin{eqnarray}
\frac{E_{\rm c1}}{\omega_0j}&=&\left\{  \begin{array}{ll}
-1& {\rm for\ }\lambda\in[\lambda_{\rm c},\lambda_0)\,,\\
-\frac{1}{2}\left(\frac{\lambda_0^2}{\lambda^2} +\frac{\lambda^2}{\lambda_0^2}\right) & {\rm for\ }\lambda\in[\lambda_0,\infty)\,,
\end{array}
\right.
\label{Ec1}
\\
&&\qquad\qquad
\lambda_0=\tfrac{\sqrt{\omega \omega_0}}{1-\delta}
\nonumber\,.
\end{eqnarray}
The energy (\ref{Ec1}) is associated with a saddle point of the energy landscape. 
Therefore, according to the classification described in Ref.\,\cite{Str16}, it corresponds to an ESQPT of type $(f,r)=(2,1)$, where $r$ is the rank of the non-degenerate stationary point (number of negative eigenvalues of the Hessian matrix).
This leads to a logarithmic divergence of the {first} derivative $d\rho/dE$ of the smoothed level density $\rho(E)$ at $E=E_{\rm c1}$ (that is a step-like but continuous behavior of $\rho$ at this energy) \cite{Str15,Str16}. 

Two other ESQPTs appear at energies \cite{Bas14,Klo17a}
\begin{eqnarray}
\frac{E_{\rm c2}}{\omega_0j}=-1 &&\quad {\rm for\ }\lambda\in[\lambda_0,\infty)
\label{Ec2}
\,,\\
\frac{E_{\rm c3}}{\omega_0j}=+1 &&\quad {\rm for\ }\lambda\in[0,\infty)
\label{Ec3}
\,.
\end{eqnarray}
They are both of the type $(f,r)=(2,2)$ and show as jumps of the first derivative $d\rho/dE$ of the smoothed level density (i.e., breaks of $\rho$) at $E_{\rm c1}$ and $E_{\rm c2}$ \cite{Str15,Str16}.

\begin{figure}[t]
\includegraphics[width=0.85\linewidth]{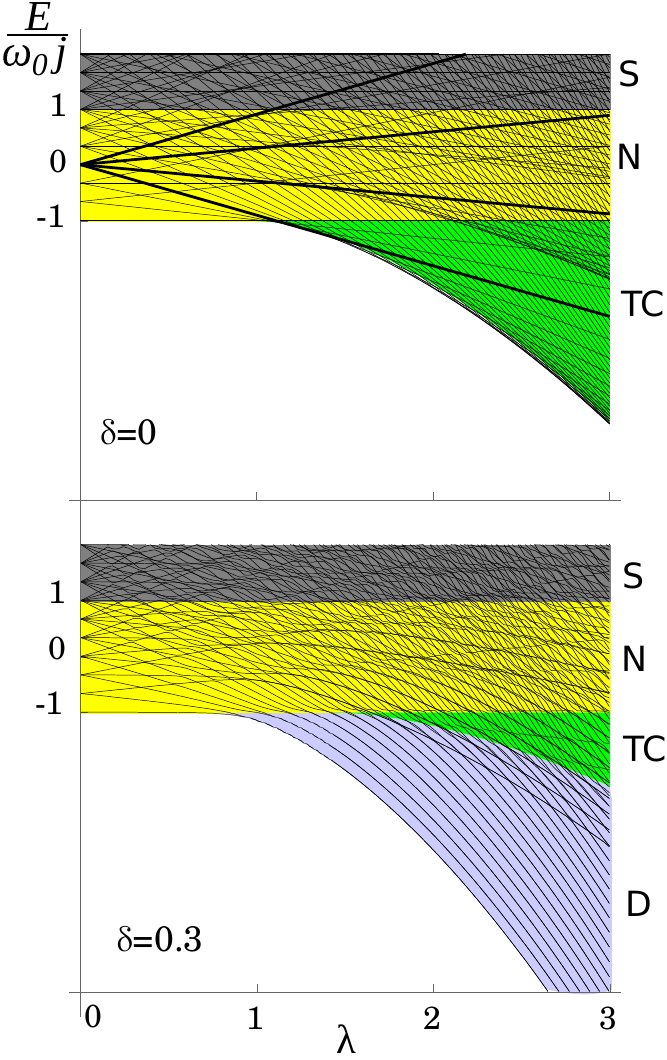}
\caption{
(Color online) 
Level dynamics and quantum phase diagrams of the extended Dicke Hamiltonian (\ref{H}) with $\delta = 0$ (upper panel)  and $\delta = 0.3$ (lower panel), both for $\omega=\omega_0=1$.
The ground-state QPTs are at $\lambda_{\rm c} = 1$ for $\delta=0$ and at $\lambda_{\rm c}\doteq 0.77$ for $\delta=0.3$. 
Quantum phases D (Dicke), TC (Tavis-Cummings), N (Normal) and S (Saturated) are marked by colors. 
ESQPT borderlines coincide with the phase boundaries.
Quantum spectra are  drawn for $j=3$.
In the $\delta=0$ panel, the $M=3$  states are plotted by thicker lines to show that levels from different $M$-subspaces mutually cross.
 }
\label{EDphase}
\end{figure}   

The diagrams showing individual energy levels  in the $\lambda\times E$ plane for a finite-$N$ realization of the $\delta=0$ and $\delta\neq 0$ models together with the $N\to\infty$ ESQPT borderlines (\ref{Ec1}), (\ref{Ec2}) and (\ref{Ec3}) are given in Fig.\,\ref{EDphase}.
The domains in between the ESQPT borderlines define quantum phases of the system.
In Fig.\,\ref{EDphase} they are marked by different colors and abbreviated as D (Dicke), TC (Tavis-Cummings), N (Normal) and S (Saturated).
The reasoning for this notation and a more detailed discussion can be found in Ref.\,\cite{Klo17a}.
Note that quantum phases cannot, in general, be distinguished by some order parameters (expectation values of suitably selected observables in individual eigenstates), but rather by different energy dependences (trends) of these expectation values smoothed over neighboring eigenstates \cite{Cej16,Klo17a}. 
 
In the Tavis-Cummings $\delta=0$ limit of the model \cite{Tav68}, the treatment of phases can be qualitatively simplified. 
In this case, the Hamiltonian (\ref{H}) has an additional integral of motion 
\begin{equation}
\hat{M}=\hat{b}^\dagger\hat{b}+\hat{J}_z+j 
\label{M}
\end{equation}
and the system is integrable [note that a general $\delta\neq 0$ Hamiltonian conserves only the parity $\hat{\Pi}\=\exp(i\pi\hat{M})$].
The value of the conserved quantity can be written as $M=n\+n^*$, where $n$ is the number of bosons and $n^*=m\+j$ the number of excited atoms. 
The total spectrum of quantum energy levels for the $\delta=0$ system with any $\lambda$ is comprised of mutually non-interacting sub-spectra with different values of $M=0,1,2,\dots$ (see the upper panel of Fig.\,\ref{EDphase}).
Each of these spectra separately can be subject (in the $N\to\infty$ limit) to a semi-classical phase transitional analysis.
To do so, it is convenient to use a canonical transformation that reduces the number of effective degrees of freedom of the $\delta=0$ system to $f=1$ \cite{Klo17a,Klo17b}.
The transformed classical Hamiltonian depends only on one pair of new conjugate variables and on the conserved quantity $M$, thus allows one to identify stationary and quasi-stationary points for different values of $M$.

\begin{figure}[t]
\includegraphics[width=0.8\linewidth]{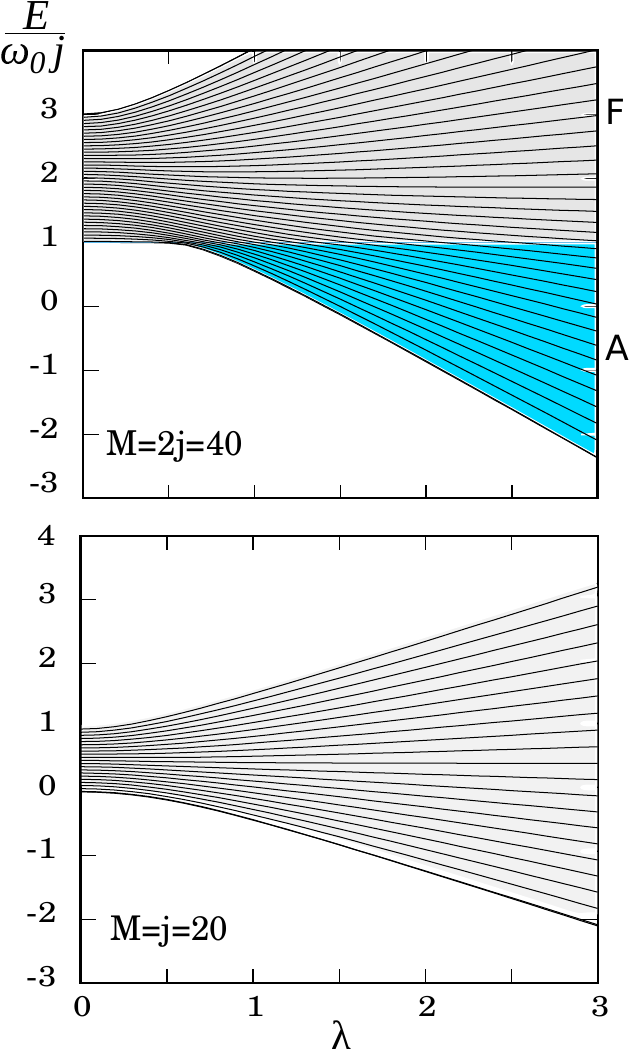}
\caption{
(Color online) 
Energy spectra of two $M$-subspaces of the $\delta=0$ model with $N=2j=40$ and $\omega_0=\omega/2=1$.
The upper panel shows a critical subspace $M=2j$ with a QPT and ESQPT. 
Quantum phases are distinguished by colors and acronyms A (Atomic) and F (Field). 
The lower panel shows a non-critical subspace $M=j$.
}
\label{TCphase}
\end{figure}   

The results of the semi-classical analysis of the $\delta=0$ model are the following \cite{Klo17a,Klo17b}:
While the subspaces with $M\neq N$ show no critical effects, the one with $M=N$ has both a QPT and an ESQPT.
Indeed, the energy of the lowest state in the $M=N$ subspace in the $N\to\infty$ limit for the $\omega>\omega_0$ hierarchy is given by
\begin{eqnarray}
\frac{E_{\rm l.s.}}{\omega_0j}=&&
\left\{\begin{array}{ll}
+1 & \lambda\leq\bar{\lambda}_{\rm c}\,,\\
+1-\frac{4}{\omega_0}g(\lambda)\left[\lambda\sqrt{1-g(\lambda)}-\bar{\lambda}_{\rm c}\right] & \lambda>\bar{\lambda}_{\rm c}\,,
\end{array}\right.
\label{Egs_TC}
\\
&& \qquad
g(\lambda)=\tfrac{2}{3}-\tfrac{2}{9}\bigl(\tfrac{\bar{\lambda}_{\rm c}}{\lambda}\bigr)^2-\tfrac{2}{9}\tfrac{\bar{\lambda}_{\rm c}}{\lambda}\sqrt{\bigl(\tfrac{\bar{\lambda}_{\rm c}}{\lambda}\bigr)^2+3}
\,,
\nonumber
\end{eqnarray}
where the critical coupling
\begin{equation}
\bar{\lambda}_{\rm c}=\frac{\omega-\omega_0}{2}
\label{LamC_TC}
\end{equation}
marks a discontinuity of $d^2E_{\rm l.s.}/d\lambda^2$, which can be interpreted as the second-order QPT in the $M=N$ subspace \cite{Fer11}.
An associated ESQPT appears at the critical energy
\begin{equation}
\frac{E_{\rm c4}}{\omega_0j}=+1 \quad {\rm for\ }\lambda\in[\bar{\lambda}_{\rm c},\infty)
\label{Ec4}
\,,
\end{equation}
where one observes divergence of the smoothed level density $\rho$ in the $M=N$ subspace.
Since the classical Hamiltonian is not analytic in this stationary point, the ESQPT classification according to Ref.\,\cite{Str16} does not work here.
Nevertheless, the observed signatures of the present ESQPT are quite similar to the case $(f,r)=(1,1)$, which is most studied in literature, see, e.g., Refs.\,\cite{Cej06,Cap08,San16,Ber17,Ley05,Rel08,Bast14,Kop15}. 

The level dynamics for two $M$-subspaces (including the critical one) of the $\delta=0$ model are shown in Fig.\,\ref{TCphase}.
In the $M=N$ subspace we indicate two quantum phases separated by the ESQPT above $\bar{\lambda}_{\rm c}$.
The phase abbreviated by A (Atomic) is characterized by a growing average $\ave{n^*}$ of the number of atomic excitations in individual eigenstates with increasing energy.
The average $\ave{n^*}$ reaches its maximum right at the ESQPT critical energy and then decreases \cite{Fer11}, which allows us to denote the quantum phase above the ESQPT by the acronym F (Field).
In this phase, the increase of energy is correlated with a growing average $\ave{n}$ of the number of bosons.

\subsection{Quantum quench dynamics}
\label{QQD}

Consider a quantum system with discrete energy spectrum described by a general Hamiltonian 
\begin{equation}
\hat{H}(\lambda)=\hat{H}_0+\lambda\hat{V}
\label{Hlin}
\end{equation}
depending linearly on a control parameter $\lambda$.
As in the case of the extended Dicke model (\ref{H}), the term $H_0$ represents a free Hamiltonian while $V$ is an interaction.
We assume $[\hat{H}_0,\hat{V}]\neq 0$ since otherwise everything would be trivial.
Suppose that the system is initially prepared in the $k$th eigenstate $\ket{\phi_k(\lambda_{\rm i})}\equiv\ket{\phi_{{\rm i}k}}\equiv\ket{\psi_{{\rm i}}}$ with energy $E_k(\lambda_{\rm i})\equiv E_{{\rm i}k}\equiv E_{\rm i}$ associated with the initial Hamiltonian $\hat{H}(\lambda_{{\rm i}})\equiv\hat{H}_{\rm i}$, and that the control parameter is suddenly changed from  $\lambda_{{\rm i}}$ to $\lambda_{{\rm f}}$.
The initial state is no more an eigenstate of the final Hamiltonian $\hat{H}(\lambda_{{\rm f}})\equiv\hat{H}_{\rm f}$ and thus undergoes a non-trivial evolution with time $t$:
\begin{equation}
\ket{\psi_{\rm f}(t)}=e^{-i\hat{H}_{\rm f}t}\ket{\psi_{\rm i}}
\label{evo}\,,
\end{equation}
where we assume $\hbar=1$.
The decay and recurrences of the initial state can be monitored by the survival amplitude $A(t)=\scal{\psi_{\rm i}}{\psi_{\rm f}(t)}$ (here and below we assume that all states are normalized).
Note that in the present setting, when the initial state is associated with a single eigenstate of the initial Hamiltonian, the survival probability $P(t)=|A(t)|^2$ is equal to the so-called  Loschmidt echo or fidelity (the probability of the initial state recovery after the forward evolution by $\hat{H}_{\rm f}$ and a backward evolution by $\hat{H}_{\rm i}$, or equivalently, the instantaneous overlap of states evolved simultaneously by $\hat{H}_{\rm i}$ and $\hat{H}_{\rm f}$) \cite{Per84a,Gor06,Gou12}.
However, this connection is broken for more general initial states.

Let us introduce the basis of the final Hamiltonian eigenvectors $\ket{\phi_l(\lambda_{\rm f})}\equiv\ket{\phi_{{\rm f}l}}$ and the corresponding set of eigenvalues $E_{{\rm f}l}$.
The distribution of the initial state in the final Hamiltonian eigenstates is expressed by the strength function (also called the local density of states)
\begin{equation}
S(E)=\sum_l \underbrace{\abs{\scal{\psi_{\rm i}}{\phi_{{\rm f}l}}}^2}_{\abs{s_l}^2} \delta(E-E_{{\rm f}l})
\label{Str}\,.
\end{equation}
It represents a probability distribution for energy after the shift $\lambda_{{\rm i}}\to\lambda_{{\rm f}}$, or shortly a distribution of final energy in the initial state.
Besides the smoothened shape of the strength function, important information is contained also in its autocorrelation function:
\begin{eqnarray}
C(E)&=&\sum_l\sum_{l'}|s_{l}|^2|s_{l'}|^2\delta(E_{{\rm f}l'}-E_{{\rm f}l}-E)
\nonumber\\
&=&\int dE'S(E')S(E'\+E)
\label{Aut}\,.
\end{eqnarray}
A trivial calculation reveals that the survival probability 
\begin{eqnarray}
P(t)&=&\biggl|\sum_l|s_l|^2e^{-iE_{{\rm f}l}t}\biggr|^2
\label{PLo1}\\
&=&\underbrace{\sum_{l}\abs{s_l}^4}_{{\cal N}^{-1}}+2\sum_l\sum_{l'(<l)}\abs{s_l}^2\abs{s_{l'}}^2\cos[(E_l\-E_{l'})t]
\nonumber
\end{eqnarray}
can be expressed via the Fourier transforms of both the strength function and its autocorrelation function:
\begin{equation}
P(t)=\abs{\int\!\!dE\,S(E)e^{-iEt}}^2=\int dE\,C(E)e^{+iEt}
\label{PLo2}.
\end{equation}
This turns out important for the interpretation of the quantum quench dynamics in various situations.
Note that the quantity ${\cal N}=1/\sum_{l}\abs{s_l}^4$, called the  participation ratio, expresses a principal number of components of the strength function \eqref{Str}. 
It varies from ${\cal N}=1$, for a perfectly localized strength function with only a single non-zero coefficient $s_l$, to ${\cal N}\to\infty$, for totally delocalized strength functions with components uniformly spread over an asymptotically increasing number of states.

The average and variance of the distribution \eqref{Str} can be determined from the relation $\hat{H}_{\rm f}\=\hat{H}_{\rm i}\+\Delta\lambda\,\hat{V}$ (where $\Delta\lambda\=\lambda_{\rm f}\-\lambda_{\rm i}$), which follows from the linearity of Hamiltonian \eqref{Hlin}.
The average is given by
\begin{eqnarray}
\ave{E_{\rm f}}_{\rm i}&=&\int\!\!dE\,S(E)E=\matr{\psi_{\rm i}}{\hat{H}_{\rm f}}{\psi_{\rm i}}
\nonumber\\
&=&E_{\rm i}+\Delta\lambda\,\ave{V}_{\rm i}
\label{AvE},
\end{eqnarray}
where $\ave{V}_{\rm i}=\matr{\psi_{\rm i}}{\hat{V}}{\psi_{\rm i}}$, while the variance reads
\begin{eqnarray}
\dis{E_{\rm f}}_{\rm i}&=&
\int\!\!dE\,S(E)\bigl(E\-\ave{E_{\rm f}}_{\rm i}\,\bigr)^2
\label{VaE}\\
&=&\matr{\psi_{\rm i}}{\hat{H}_{\rm f}^2}{\psi_{\rm i}}\-\matr{\psi_{\rm i}}{\hat{H}_{\rm f}}{\psi_{\rm i}}^2=(\Delta\lambda)^2\dis{V}_{\rm i}
\nonumber,
\end{eqnarray}
where $\dis{V}_{\rm i}\=\matr{\psi_{\rm i}}{\hat{V}^2}{\psi_{\rm i}}\-\matr{\psi_{\rm i}}{\hat{V}}{\psi_{\rm i}}^2$.
Due to the Hellmann-Feynman formula  $\ave{V}_{\rm i}\=dE_{\rm i}/d\lambda$, the relation \eqref{AvE} can be used to determine the final energy average, i.e., a centroid of the distribution \eqref{Str}, from the position and tangent of the selected energy level at the initial parameter value.
This allows one to design specific quench protocols that probe selected parts of the spectrum of the final Hamiltonian, for example, different quantum phases of the system and various ESQPT critical domains \cite{Fer11}.
However, according to \Eqref{VaE}, the final energy variance, i.e., squared width of the distribution \eqref{Str}, is proportional to the variance of $V$ in the initial state and grows with the square of $\Delta\lambda$.
This sets unavoidable limits to the probing procedure since the dispersion of the final energy distribution implies averaging of the response over a broader interval of the spectrum, hence reduces the resolution of the procedure.

The evolution of the survival probability on various time scales defines different regimes of the quench dynamics \cite{Bor16,Tav17,Tor18,Gor06,Gou12,Tav17}.
They are governed by physical mechanisms that naturally follow from an increasing energy resolution with which the strength function \eqref{Str} is being reflected by the evolving system at the given instant of time.
The regimes of quantum quench dynamics can be schematically described as follows:

(a) Ultra-short time regime, $t\ll t_{\rm s}$, where
\begin{equation}
t_{\rm s}=\frac{1}{\sqrt{\dis{E_{\rm f}}_{\rm i}}}
\label{ts}
\end{equation}
is the time derived from the final energy dispersion \eqref{VaE}:
At this time scale, the system can feel merely the width of the strength function and decays according to the simple quadratic formula $P(t)\approx 1-(t/t_{\rm s})^2+\dots$.
This stage of evolution carries no information on the final Hamiltonian.

(b) Short- and medium-time regime, from $t\sim t_{\rm s}$ up to times $t\ll t_{\rm H}$ well before the Heisenberg scale set by Eq.\,\eqref{Heit} below:
In this regime, the energy resolution becomes sufficient to distinguish an outline shape of the strength function \eqref{Str} as well as some of its correlation properties given by \Eqref{Aut}.
Qualified estimates of the shape in various situations predict an initially exponential, Gaussian or sub-Gaussian decrease of the survival probability \cite{Bor16,Tav17}.
The first dip of $P(t)$ (a \uvo{survival collapse}) is sometimes followed by modulated oscillations with a power-law decrease of their amplitude (related for instance to low- and/or high-energy edges of the strength function) \cite{Tav16,Tav17}.

(c) Long-time regime, around $t\sim t_{\rm H}$:
The Heisenberg time is computed according to
\begin{equation}
t_{\rm H}=\frac{2\pi}{\sum_l\frac{1}{2}\bigl[|s_{l+1}|^2\+|s_l|^2\bigr]\bigl[E_{{\rm f}(l+1)}\-E_{{\rm f}l}\bigr]}=\frac{2\pi}{\ave{\Delta E_{\rm f}}_{\rm i}}
\label{Heit}\,,
\end{equation}
where $\ave{\Delta E_{\rm f}}_{\rm i}$ is an average spacing of the final energy levels in the initial state distribution.
At this time scale, the system gradually resolves the discrete structure of the strength function, from smaller to larger level density domains.
Power-law modulated oscillations can appear also at this stage, being connected with the behavior of the autocorrelation function for small energy differences \cite{Tav17,Ler18}. 
They may be followed by a so-called  correlation hole---a long-lasting suppression of the survival probability below its asymptotic-time average, which reflects strong correlations of individual levels in chaotic systems \cite{Bor16,Tor18}.

In Sec.\,\ref{NUR}, we will encounter situations in which the strength function populates considerably only a certain subset of states of the final Hamiltonian.
In these cases it is convenient to introduce a modified  Heisenberg time $t'_{\rm H}$ that takes the partial fragmentation into account.
It is computed in the same way as the standard Heisenberg time $t_{\rm H}$ in Eq.\,\eqref{Heit}, but only with a reduced set of levels $E_{{\rm f}l}$ obtained by removing the states with the lowest values of $|s_l|^2$.
In the numerical calculations below we select a threshold for the state removal given by $0.5 \%$ of the total strength.
For partially fragmented states, $t'_{\rm H}$ gives a better prediction on where the discrete structure of the strength function starts to play a role in the quench dynamics.
If the strength function is fully fragmented, $t_{\rm H}$ and $t'_{\rm H}$ tend to coincide.

(d) Ultra-long  time regime, $t\gg t_{\rm H}$:
The infinite-time average and variance of the function $P(t)$ in \Eqref{PLo1} read
\begin{eqnarray}
\overline{P(t)}&=&{\cal N}^{-1}
\label{Pave}\,,\\
\overline{P(t)^2}-\overline{P(t)}^2&=&{\cal N}^{-2}-\sum_l|s_l|^8
\label{Pvar}\,,
\end{eqnarray}
where bars represent time averaging of the respective quantities according to $\overline{g(t)}\=\lim_{T\to\infty}\int_0^Tg(t)dt/T$.
So in the very long time perspective, the survival probability can be seen as fluctuations around the \uvo{saturation value} \eqref{Pave} with standard deviation given by the the square root of \eqref{Pvar}.
Both these quantities decrease with the degree of fragmentation of the corresponding strength function \eqref{Str}.
Note that for strongly delocalized states, the second term on the right-hand side of \Eqref{Pvar} gives a contribution $\sim{\cal N}^{-3}$, which is negligible relative to the first term, while for localized states this term causes a considerable reduction of the variance.

Despite a usually low average \eqref{Pave}, the ultra-long time regime unavoidably includes also sharp peaks of $P(t)$ reaching values even very close to unity.
These partial revivals of the initial state demonstrate the well-known quantum recurrence theorem \cite{Boc57}, which guarantees that for any initial state of a system with discrete spectrum and for an arbitrary degree of precision there exists a time at which the evolved state restores the initial one with this precision.
As follows from \Eqref{Pvar}, a higher frequency of recurrences is expected for less fragmented strength functions and vice versa.

A valuable  insight into the survival probability evolution can be gained from the quasi-classical picture of quantum dynamics.
Associating with the state $\ket{\psi_{\rm f}(t)}$ at any stage of its evolution the Wigner phase-space distribution function $W(q,p,t)$, we can rewrite the survival probability as
\begin{equation}
P(t)=2\pi\iint dq\,dp\,W(q,p,t)W(q,p,0)
\label{PLoc}\,,
\end{equation}
where $q$ and $p$ stand for $f$-dimensional vectors of mutually conjugate coordinates and momenta, respectively.
Assume that $W(q,p,t)$ is classical-like (i.e., shows only negligible domains with negative values) or is transformed to such form by a convenient smoothing procedure $W(q,p,t)\to\overline{W}(q,p,t)$.
Then the evolution can be approximated by means of the equations of motions derived from the classical Hamiltonian function $H_{\rm f}(q,p)$ corresponding to $\hat{H}_{\rm f}$.

The classical treatment of the smooth(ed) Wigner function $\overline{W}(q,p,t)$ and its evolution allows one to estimate possible signatures of classical stationary points in the survival probability $P(t)$, and therefore to partly anticipate an influence of QPTs and ESQPTs on the quench dynamics.
Consider a stationary point $(q_{\rm s},p_{\rm s})$ of the function $H_{\rm f}$ at energy $E_{\rm s}=H_{\rm f}(q_{\rm s},p_{\rm s})$.
If $E_{\rm s}$ belongs to the support of a smoothed strength function $\overline{S}(E)$, some effects of the stationary point may be seen in $P(t)$ for $t\lesssim t_{\rm H}$.
The form of these effects is expected to depend on whether the stationary point $(q_{\rm s},p_{\rm s})$  is located within the phase-space domain where the initial distribution $\overline{W}(q,p,0)$ yields considerable contributions, or whether the stationary point is outside that domain.
In the first case, the decay of the survival probability \eqref{PLoc} gets slowed down at its initial stage, $t\lesssim t_{\rm s}$, due to the slow classical dynamics around $(q_{\rm s},p_{\rm s})$.
A clear demonstration of this behavior within the extended Dicke model will be presented in Secs.\,\ref{FQF1} and\,\ref{FQPF2}.

On the other hand, if the stationary point is located outside the domain with large values of $\overline{W}(q,p,0)$, the short-time decay of $P(t)$ remains unaffected.
Nevertheless, an indirect effect may be observed at some later stages of the $P(t)$ evolution, when the stationary point $(q_{\rm s},p_{\rm s})$ prevents the return of a certain fraction of the $\overline{W}(q,p,t)$ distribution (that with energy close to $E_{\rm s}$) back to the initial phase-space domain.
Then we may expect a partial reduction of the survival probability $P(t)$ for times comparable with the Heisenberg scale, $t\sim t_{\rm H}$, which coincides with an average classical return time.
Indications of such behavior will be indeed discussed in Secs.\,\ref{BQF1} and\,\ref{BQPF2}, but we stress here that the reduction size (the possibility to actually observe any effect) strongly depends on the degree of stability (chaos) of classical motions generated by $H_{\rm f}(q,p)$ in the relevant phase-space domain.

\section{Numerical results}
\label{NUR}

In this section, numerical results on the quantum quench dynamics in the extended Dicke model with Hamiltonian \eqref{H} will be analyzed.
Subsection\,\ref{F1} deals with the quenches in $M$-subspaces of the integrable $\delta=0$ (Tavis-Cummings) regime where the dynamics is effectively reduced to one degree of freedom.
Subsection\,\ref{F2}  is focused on the quenches in the full $\delta \neq 0$ model with two degrees of freedom.
\subsection{Integrable $\delta=0$ regime }
\label{F1}

\begin{figure}[t]
\includegraphics[width=1\linewidth]{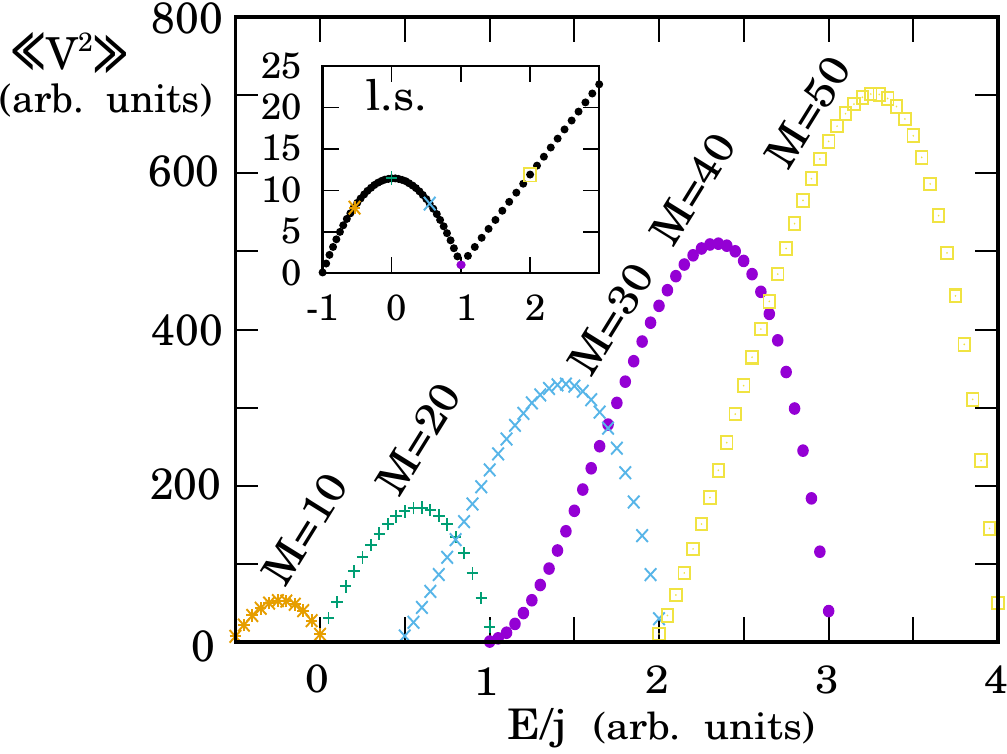}
\caption{
(Color online) Dispersion $\dis{V}_{\rm i}$ from Eq.\,\eqref{Eq:DispAnal} in the unperturbed eigenstates $|\psi_{\rm i}\rangle = \ket{n}\ket{m}$ of the $\delta=0$ model plotted against their energies.
Parameters of the system are $\omega=2, \omega_0=1$, $j=20$.
The inset shows $\dis{V}_{\rm i}$ in the lowest states from all $M$-subspaces.
The states involved in the main panel are marked in the inset by the respective symbols.
}
\label{FDispersion}
\end{figure} 

\begin{figure*}[t]
\includegraphics[width=0.95\linewidth]{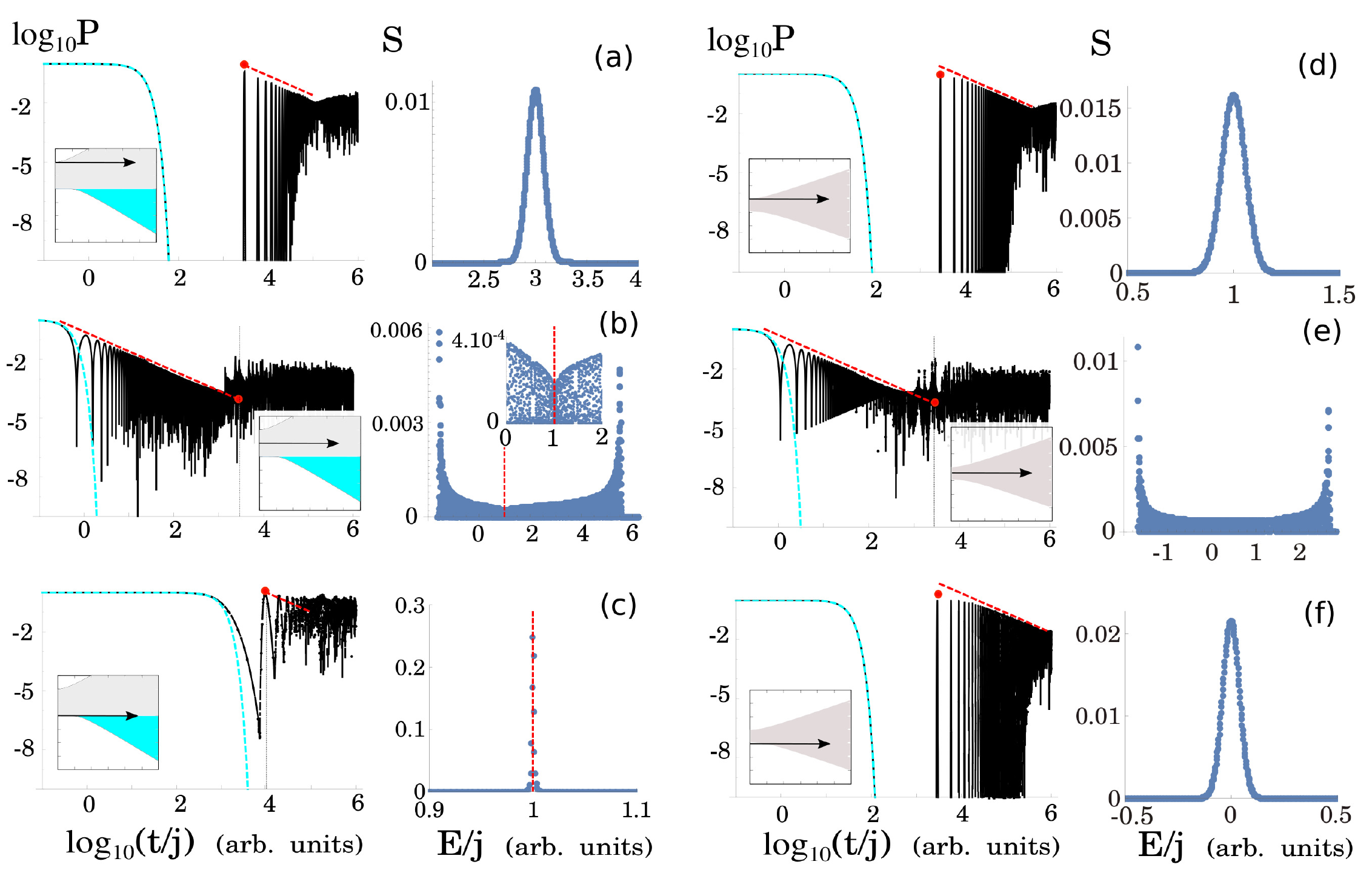}
\caption{
(Color online) The survival probability and the strength function corresponding to the  FQPs of the $\delta=0$ model with $j=2000$ in off-resonant setting $\omega=2, \omega_0=1$. 
The critical  $M=2j$  subspace is shown in panels (a)--(c), a non-critical $M=j$ subspace in panels (d)--(f). 
In all the cases, $\lambda_{\rm i}=0$ and $\lambda_{\rm f}=2.5$ (the quench protocol is visualized by an arrow in the respective  phase diagram).
The real decay (black curves) is compared with the Gaussian decay (the light-blue dashed curves), the Heisenberg time  $t_{\rm H}$ is marked with the red bullets, and the $ 1/t$  decay of oscillations   is marked with the red dashed lines.
In the strength function plots, the position of the ESQPT is indicated by vertical lines.
}
\label{FQPTC}
\end{figure*}

\subsubsection{Forward quench protocols, $f=1$}
\label{FQF1}

The evolution of the survival probability strongly depends on the quench protocol, that is on the selection of the initial state and on the size of the parameter change.
In the {forward quench protocols} (FQPs) we set initial states as various eigenstates of the unperturbed $\lambda_{\rm i}\=0$ Hamiltonian and choose the final value $\lambda_{\rm f}>0$.

The decay rate at ultra-short and short times of such initial states can be  estimated  using Eqs.\,\eqref{VaE} and\,\eqref{ts}.
In the detuned system $\omega \neq \omega_0$ (the initial eigenstates are non-degenerate, hence $|\psi_{\rm i}\rangle=\ket{n}\ket{m}$), a simple formula for the dispersion of the interaction Hamiltonian term can be obtained:
\begin{equation}
\dis{V}_{\rm i}=\frac{(1+\delta^2)(j^2-m^2+j)(2n+1)+(1-\delta^2)m}{2j}. 
 \label{Eq:DispAnal}
\end{equation}
In Fig.\,\ref{FDispersion} we show  $\dis{V}_{\rm i}$ in multiple eigenstates belonging to several $M$-subspaces for $j=20$.
For all the subspaces we observe a similar dependence---the states closer to the edges of the spectrum have smaller dispersion than the ones in the middle and therefore their decay is slower.
However, a closer look reveals an anomaly for the critical subspace $M=2j=40$.
The inset of Fig.\,\ref{FDispersion} depicts the dispersion of the lowest state from all subspaces with $M=0,1,2 \dots$,  plotted against their energies $E_{\rm l.s.}(M)=\omega(M-j)-(\omega-\omega_0) j$.
The leftmost point corresponding to the global ground state has $\dis{V}_{\rm i}=0$.
Indeed, as it is the only member of the $M=0$ subspace  it cannot decay.
However, small values of dispersion are reached also for the $M$-subspaces close to the critical one with $\dis{V}_{\rm i}=1/2j$, which indicates an asymptotically slow decay of the respective initial states in the $j\to\infty$ limit.

Let us now proceed to concrete examples of FQPs within two $M$-subspaces, the critical one with $M=2j$ and the non-critical with $M=j$.
In the following we consider $j=2000$.
Fig. \ref{FQPTC} depicts both the survival probability and the strength function for several initial states from the above mentioned subspaces.

In the first row of Fig.\,\ref{FQPTC} (panels\,a and\,d) we compare the highest excited states.
The envelope of the strength function has a Gaussian shape, giving rise to an initial Gaussian decay of the survival probability \cite{Tav17}.
After the initial decay, strong revivals appear at about the Heisenberg time $t_{\rm H}$.
Their amplitude decreases as $\propto 1/t$ until the saturation regime  around $P(t) \sim \mathcal{N}^{-1}$ is reached.
The power-law modulation $\propto 1/t^\alpha$  of the oscillations with various exponents $\alpha$ was  observed in various systems and has been attributed to several specific mechanisms\,\cite{Tav16,Tav17,Ler18}.
The present case $\alpha=1$ results from two conditions: an approximately Gaussian envelope of the strength function and its discrete energy sampling 
\begin{equation}
E_{{\rm f}l}\approx e_0+e_1l+e_2 l^2
\label{Eq:scaling}
\end{equation} 
with parameters $e_0, e_1$ and $e_2\neq 0$ satisfying $|e_2| \ll |e_1|$\,\cite{Ler18}.
As can be numerically checked, both these conditions are valid in our case.

In the second row of Fig. \ref{FQPTC} (panels\,b and\,e) we compare the decay of $\lambda_{\rm i}=0$ initial states from  the middle of $M=2j$ and $M=j$  spectra.
In both cases, the strength function has a bimodal shape with large dispersion.
As a result, the initial decay is  faster than Gaussian. 
We again observe $\propto 1/t$ modulated oscillations, but \textit{before} the Heisenberg time $t_{\rm H}$.
In this case, the origin of the power-law dependence lies purely in the profile of the strength function, namely in its U-shaped envelope.
Although the ESQPT does not visibly affect the survival probability, the inset of panel\,(b) shows that the strength function forms a small dip at the critical energy.

Finally, the last row of Fig. \ref{FQPTC} (panels c and f) depicts FQPs with the lowest states from both $M=2j$ and $M=j$  subspaces.
Panel\,(c) shows the critical quench---the initial ground state is displaced directly into the region of ESQPT between the A and F phases at energy $E_{{\rm c}4}$ (see the phase diagram inset).
The initial decay is significantly slowed down (even slower than the Gaussian decay).
Semiclassically this can be viewed as a slowdown of the dynamics  due to the localization of the initial state at the stationary point $(q_{\rm s},p_{\rm s})$ of the final Hamiltonian, see the end of Sec. \ref{QQD}.
A very narrow strength function indicates a high localization of the initial state  in the final eigenbasis.
As the Gaussian envelope is lost, we do not observe any power-law modulated oscillations around the Heisenberg time.
On the other hand, in the non-critical subspace  (panel\,f) we obtain a similar decay pattern as for the highest excited state (panel\,d), manifesting that the presence of an ESQPT is crucial for the existence of the localization.
The stabilization of the initial state due to an ESQPT within a similar quench protocols in different $f=1$ systems  was also studied in Refs.\,\cite{Tor18,Ber17}.

\subsubsection{Backward quench protocols, $f=1$}\label{BQF1}

\begin{figure}[t]
\includegraphics[width=0.9\linewidth]{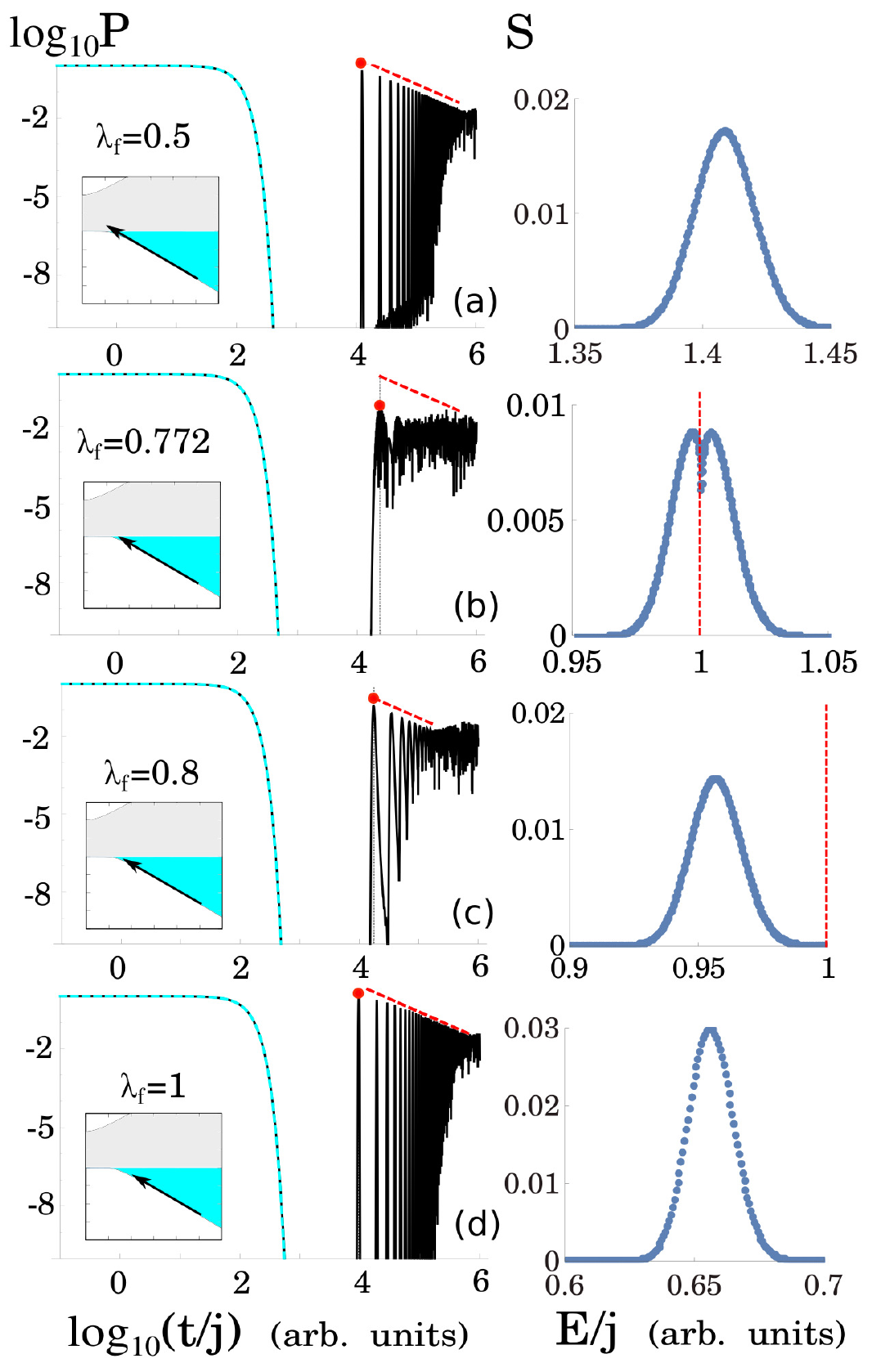}
\caption{
(Color online) The survival probability and the strength function corresponding to the  BQPs in the critical  $M=2j$ subspace of the $\delta=0$ model.
The parameters and  meaning of symbols are the same as in Fig.\,\ref{FQPTC}.
}
\label{FBPTC}
\end{figure} 

\begin{figure}[t]
\includegraphics[width=\linewidth]{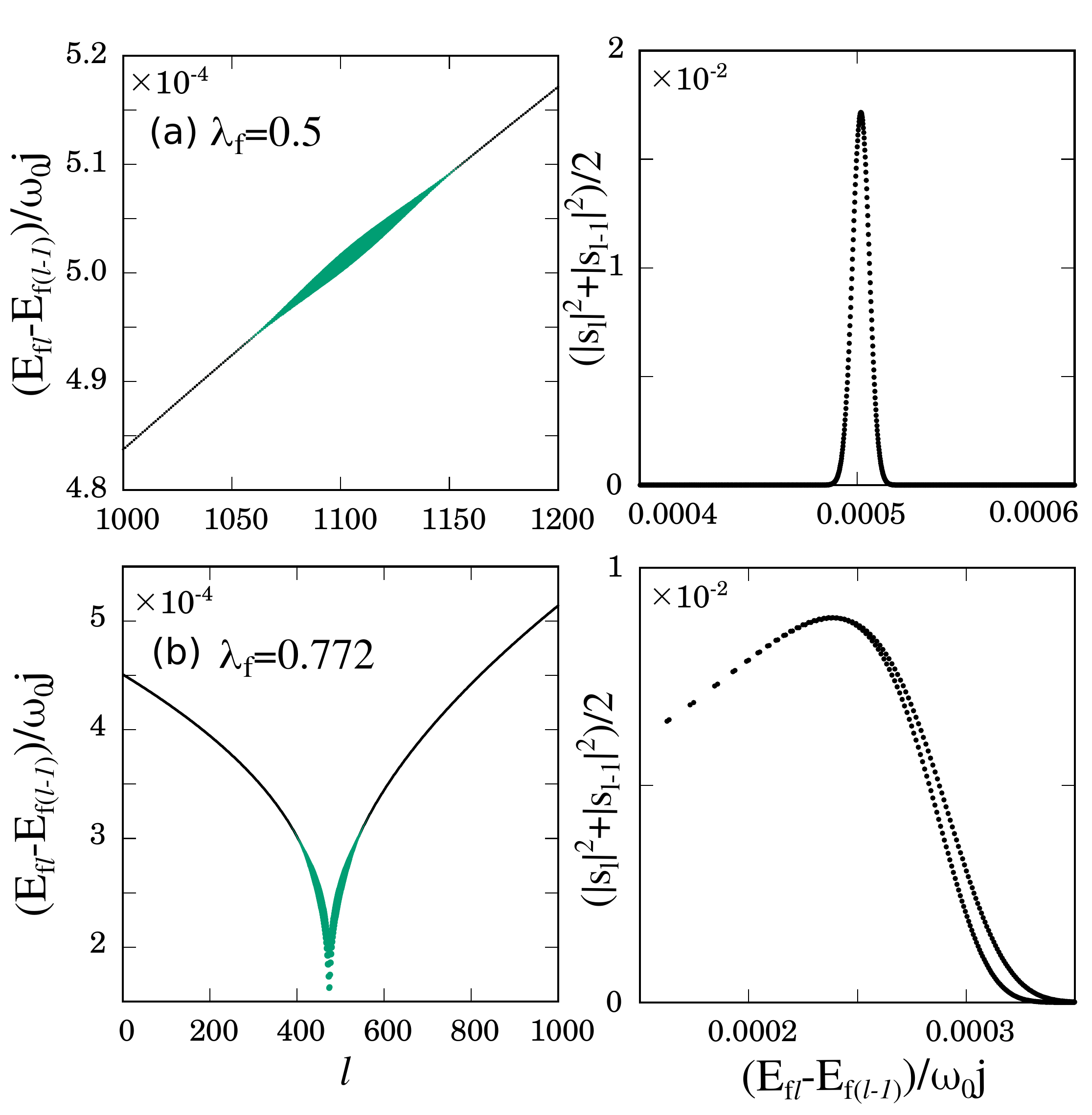}
\caption{
(Color online)  Distributions of the the energy spacing between neighboring levels of the final Hamiltonian after the quenches from panels (a) and (b) of Fig.\,\ref{FBPTC}.
Left graphs show the dependence of the energy spacing on the level index $l$. 
The green dots mark the levels populated in the quench---the dot area is proportional to the mean size of the strength function in the respective pair of levels.
The plots on the right display the level spacings correlated with the mean sizes of the strength function.
The critical quench in the lower row yields dependences with two distinct branches.
}
\label{novak}
\end{figure} 

In the backward quench protocols (BQPs), we set $\lambda_{\rm i}$ above the critical value (in this case $\bar{\lambda}_c$) and  choose various values of $\lambda_{\rm f}<\lambda_{\rm i}$ \cite{Fer11} .
In Fig.\,\ref{FBPTC} we consider the initial ground state at $\lambda_{\rm i}=2.5$.
The survival probabilities and strength functions for $\lambda_{\rm f}=0.5$ (panel\,a), $\lambda_{\rm f}=0.8$ (panel\,c) and $\lambda_{\rm f}=1$ (panel\,d) are qualitatively similar to those in panels (a), (d) and (f) of Fig.\,\ref{FQPTC}.
However, the quench with $\lambda_{\rm f}=0.772$ in panel (b) of Fig.\,\ref{FBPTC} has a different character.

The quench in Fig.\,\ref{FBPTC}(b) is critical in the sense that its final state population is centered roughly at the ESQPT energy $E_{{\rm c}4}$.
We see that the corresponding strength function has a bimodal form with a dip at the critical energy. 
Note that a similar behavior [see also Fig.\,\ref{FQPTC}(b)] would be observed for quenches within a certain interval around the present value of $\lambda_{\rm f}$. 
The initial decay of the survival probability after the critical quench does not differ from the other cases in Fig.\,\ref{FBPTC}, but the $\propto 1/t$ dependence of revivals after the Heisenberg time is not present.
The evolution of the survival probability gets to the saturation regime right after the survival collapse, which can be interpreted as a speed-up of the decay.
This difference from the FQP case, where the ESQPT caused a longer survival, demonstrates that the quench protocol (the choice of the initial state) plays an important role for the ESQPT-induced effects.

The reason for absence of the $1/t$ behavior lies in the violation of formula \eqref{Eq:scaling} for quenches populating states across the ESQPT. 
This is demonstrated in Fig.\,\ref{novak}, where correlations between the energy spacings $E_{{\rm f}l}-E_{{\rm f}(l-1)}$ and the mean populations $(|s_l|^2+|s_{l-1}|^2)/2$ of the respective neighboring levels are visualized for quenches from panels (a) and (b) of Fig.\,\ref{FBPTC}.
The left column in Fig.\,\ref{novak} shows the energy spacing as a function of $l$, with the mean populations marked by sizes of the green dots. 
The right column depicts the energy spacing versus the mean population.
In the upper row of Fig.\,\ref{novak}, which corresponds to the non-critical quench, we see that the energy spacing is approximately a linear function of $l$, in agreement with Eq.\,\eqref{Eq:scaling}, which leads to a sharply peaked distribution of energy spacings in the populated ensemble of levels.
In contrast, for the critical quench in the lower row both dependences exhibit two distinct branches.
These are associated with the states below and above the critical energy $E_{{\rm c}4}$, where the spacing vanishes. 

From the semiclassical viewpoint, the suppression of the power-law oscillations for the critical quench in Fig.\,\ref{FBPTC}(b) can be attributed to some peculiar features of the long-time dynamics of the phase-space distribution associated with the evolving quantum state---see the discussion at the end of Sec.\,\ref{QQD}.
As the global minimum $(q_{0},p_{0})$ of the initial Hamiltonian $H_{\rm i}(q,p)$ is far from the $E=E_{{\rm c}4}$ stationary point $(q_{\rm s},p_{\rm s})$ of the final Hamiltonian $H_{\rm f}(q,p)$, the support of the initial state's Wigner function localized around $(q_0,p_0)$ does not considerably overlap with $(q_{\rm s},p_{\rm s})$. 
Therefore, the latter stationary point does not affect the short time decay of the initial state but only its recurrences  at the $t \sim t_{\rm H}$ time scale.

\begin{figure}[t]
\includegraphics[width=1\linewidth]{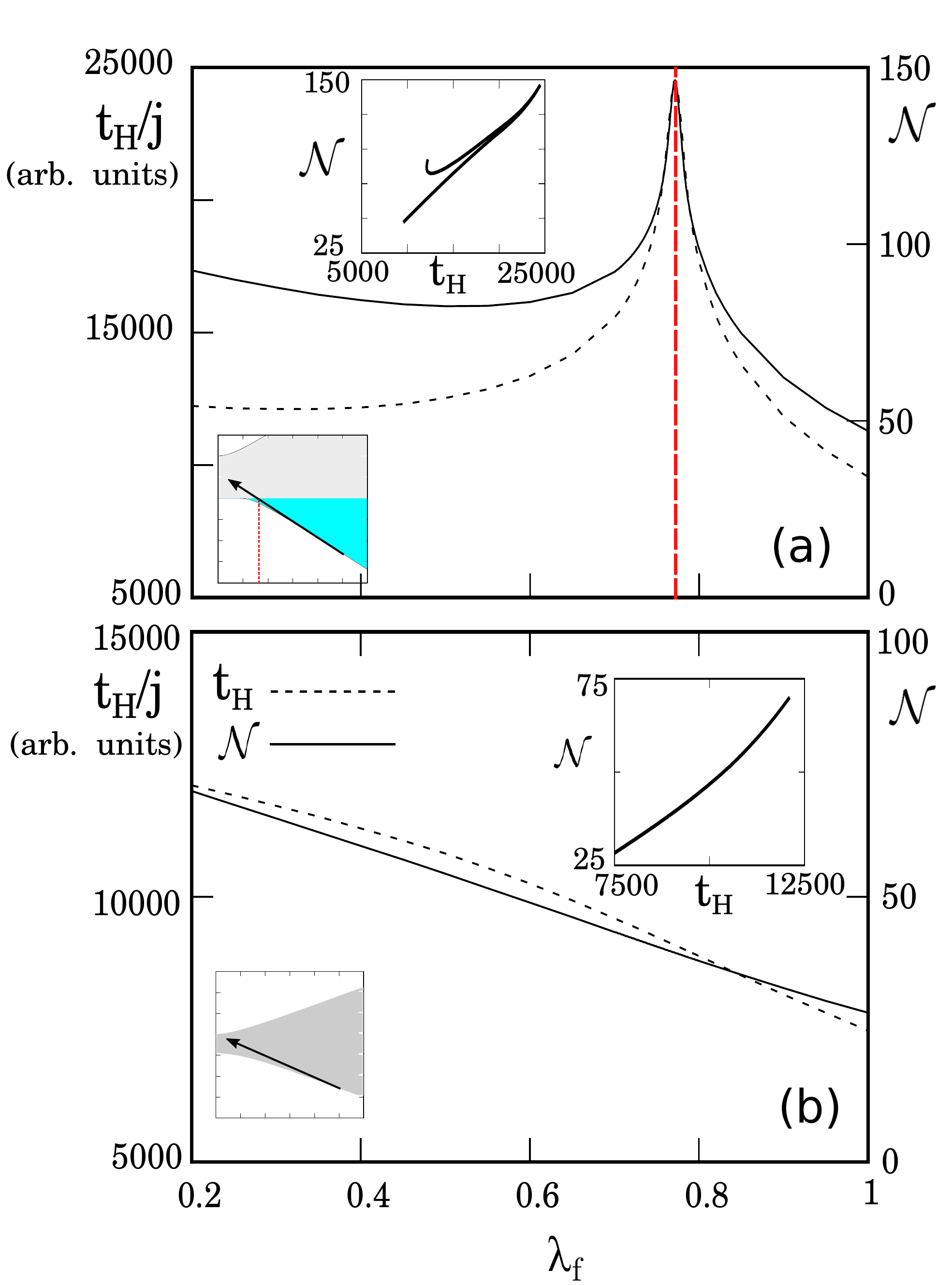}
\caption{ 
(Color online) The Heisenberg time $t_{\rm H}$ and participation ratio $\mathcal{N}$ for BQPs from $\lambda_{\rm i}=2.5$ to various values of $\lambda_{\rm f}$ in the critical  $M=2j$ (panel a) and non-critical $M=j$ (panel b) subspaces of the $\delta=0$ model.
The parameters are the same as in Fig.\,\ref{FQPTC}.
The critical quench in panel (a) is marked with vertical line.
The insets above the curves show the correlation between the quantities $t_{\rm H}$ and $\mathcal{N}$.
}
\label{F:SmoothBack}
\end{figure}
\label{BQF1}

In Fig.\,\ref{F:SmoothBack} we compare the BQPs in both critical and non-critical subspaces by plotting the Heisenberg time $t_{\rm H}$ and the participation ratio $\mathcal{N}$ as a function of $\lambda_{\rm f}$.
We see that both $t_{\rm H}$ and $\mathcal{N}$ have a sharp maximum at the critical quench in panel (a) while the non-critical dependences in panel (b) are smooth and monotonous.
This can be qualitatively understood as follows:
Given a smoothed strength function $\overline{S}(E)$ and smoothed density of states $\overline{\rho}_{\rm f}(E)$ of the final Hamiltonian (where smoothing means elimination of $\delta$-functions by a local averaging), the inverse participation ratio can be approximated by 
\begin{equation}
{\cal N}^{-1}\approx\int dE\,\overline{\rho}_{\rm f}(E)^{-1}\overline{S}(E)^2
\label{parat}\,.
\end{equation}
Assuming now (i) a Gaussian shape of $\overline{S}(E)$ with an average $\ave{E_{\rm f}}_{\rm i}$ and variance $\dis{E_{\rm f}}_{\rm i}$, and (ii) an analytic energy dependence of the inverse level density $\overline{\rho}_{\rm f}(E)^{-1}=d_0+d_1 E+d_2 E^2+\dots$ (where $d_0,d_1,d_2,\dots$ are some coefficients), we obtain the formula
\begin{equation}
{\cal N}^{-1}\approx\frac{d_0+d_1\ave{E_{\rm f}}_{\rm i}+d_2\dis{E_{\rm f}}_{\rm i}/2+\dots}{2\sqrt{\pi\dis{E_{\rm f}}_{\rm i}}}
\label{parat2}\,,
\end{equation}
which can be further transformed to the form depending on the parameter shift $\Delta\lambda$ by inserting expressions from Eqs.\,\eqref{AvE} and \eqref{VaE}.
If $|d_0|\gg|d_1|,|d_2|,\dots$, the participation ratio ${\cal N}$ becomes roughly proportional to $|\Delta\lambda|$, which is exactly the behavior observed in panel (b) of Fig.\,\ref{F:SmoothBack}.
On the other hand, the critical dependence of ${\cal N}$ shown in panel (a) is a consequence of the violation of both the above conditions (i) and (ii) for the quenches populating states across the ESQPT.
Note also that the divergence of the Heisenberg time $t_{\rm H}$ at the critical $\lambda_{\rm f}$ can be deduced from the dependences in Fig.\,\ref{novak}(b).

\subsection{General $\delta \neq 0$ regime}
\label{F2}
 \subsubsection{Forward quench protocols, $f=2$} 
 \label{FQPF2}
 
\begin{figure}[t!]
\includegraphics[width=1\linewidth]{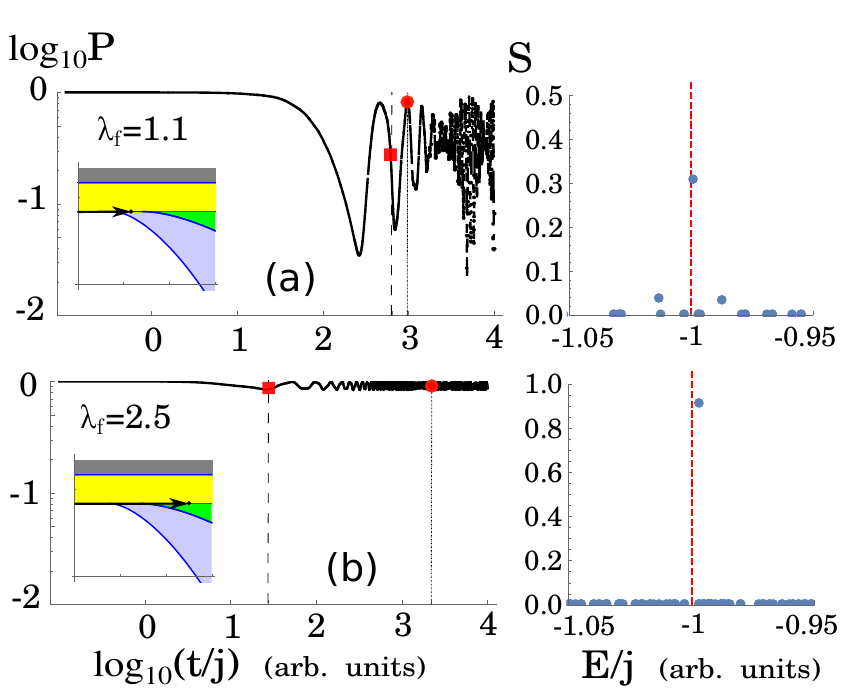}
\caption{ (Color online) The survival probability and the strength function corresponding to the FQPs from the  $\lambda_{\rm i}=0$  ground state of the $\delta=0.3$ model with parameters $j=20$, $\omega=\omega_0=1$ in the even parity sector.
The standard and modified Heisenberg times $t_{\rm H}$ and $t'_{\rm H}$ are marked with the red bullet and square, respectively.
}
\label{F:0p3FQPgs}
\end{figure}

 We now proceed to study the quench dynamics in a general $\delta \neq 0$ model with two degrees of freedom.
 We set the model parameter $\delta=0.3$ and tune the system to resonance $\omega=\omega_0=1$.
 In contrast to the integrable case, we consider only the initial states associated with the ground state of $H_{\rm i}$.
  In the FQPs, we choose the ground state at $\lambda_{\rm i}=0$ and perform a quench to $\lambda_{\rm f}=1.1$ and  $\lambda_{\rm f}=2.5$, see Fig.\,\ref{F:0p3FQPgs}.
 These values were selected because we want to test different types of ESQPTs (see the insets in the respective figure). 
 Indeed, for $\lambda_{\rm f}=1.1$ the strength function is centered at the  $(f,r)=(2,1)$ ESQPT critical energy $E_{c1}$ between the D and N phases.
 On the other hand, for $\lambda_{\rm f}=2.5$ the strength function is localized at the $(f,r)=(2,2)$ ESQPT critical energy $E_{c2}$ between the TC and N phases.
 
Panel\,(a) of Fig.\,\ref{F:0p3FQPgs} presents a similar decay pattern as the integrable case in Fig.\,\ref{FQPTC}(c).
We again observe that the strength function has only a few non-zero components in the vicinity of the ESQPT energy, indicating a rather high level of localization of the initial state in the final eigenbasis.
However, if we increase the final parameter value  to $\lambda_{\rm f}=2.5$, the localization becomes nearly perfect, see Fig.\,\ref{F:0p3FQPgs}(b). 
Indeed, the $\lambda_{\rm i}=0$ ground state has a $96\% $ overlap with the $\lambda_{\rm f}=2.5$ eigenstate closest to the ESQPT.
This difference between D-N and TC-N phase borderlines has been pointed out in Ref.\,\cite{Klo17a}. 
As a consequence, in the long-time regime the survival probability in Fig.\,\ref{F:0p3FQPgs}(b) oscillates around a quite high value $P \approx 0.85$. 
Note that the onset of oscillations neatly coincides with the modified Heisenberg time $t_{\rm H}'$ (see Sec.\,\ref{QQD}), which is marked with the red square and the vertical dashed line.

\subsubsection{Backward quench protocols, $f=2$} 
 \label{BQPF2}

\begin{figure}[t]
\includegraphics[width=1\linewidth]{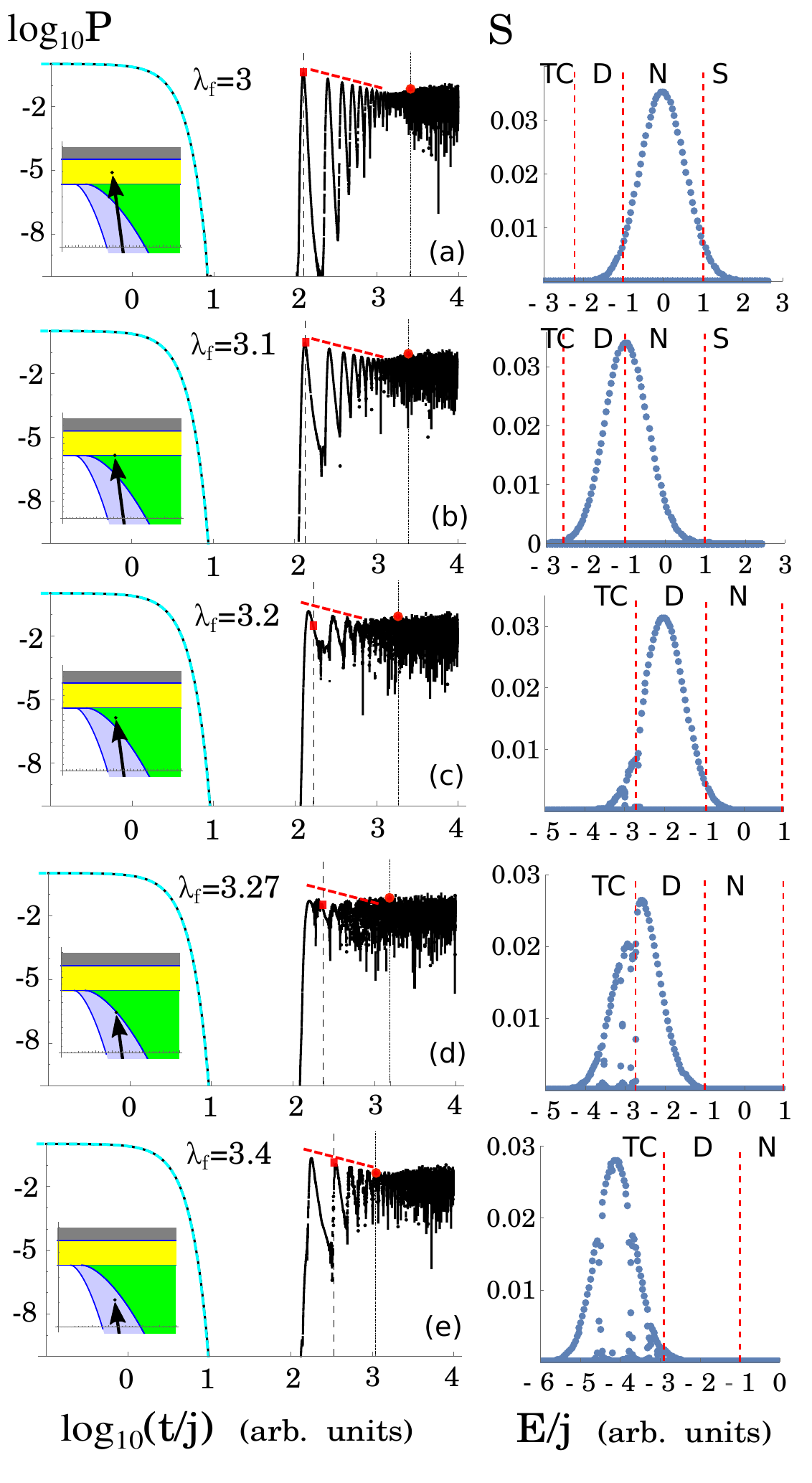}
\caption{ (Color online) The survival probability and the strength function corresponding to the BQPs from the $\lambda_{\rm i}=6$ ground state of the $\delta=0.3$ model with parameters $j=20$, $\omega=\omega_0=1$ in the even parity sector.
The standard and modified Heisenberg times $t_{\rm H}$ and $t'_{\rm H}$ are depicted with the red bullet and square, respectively.
The $1/t$ decay of oscillations is marked with the red dashed lines.
}
\label{F:0p3BQPgs}
\end{figure}

 Using the same setting of the model, we employ  BQPs starting from the superradiant ground state at $\lambda_{\rm i}=6>\lambda_{\rm c}$.
 There are several ESQPTs to be probed in this way.
 In Fig.\,\ref{F:0p3BQPgs}, results for several values of $\lambda_{\rm f}$ are depicted.
 We observe an initial Gaussian decay in all cases.
 Further, we can see that the first revival appears roughly around the modified Heisenberg time $t'_{\rm H}$.
 These revivals decay in most cases as $1/t$. 
Apparently, this behavior of the revivals is also present in the $\lambda_{\rm f}=3.1$ critical quench probing the $(f,r)=(2,2)$ ESQPT between the TC and N phases at $E=E_{{\rm c}2}$  (the same type of ESQPT between the N and S phases at $E=E_{{\rm c}3}$ was also examined, showing the same result).
 However, if we choose $\lambda_{\rm f} \doteq 3.27$, corresponding to critical quench to the $(f,r)=(2,1)$ ESQPT between the D and TC phases at $E=E_{{\rm c}1}$, we observe  the vanishing of the $1/t$ modulated revivals.
 This is again due to the splitting of the strength function at the critical energy---a similar effect as in the $f=1$ critical case, compare Fig.\,\ref{F:0p3BQPgs}(d) with Fig.\,\ref{FBPTC}(b).
  
\begin{figure}[t]
\includegraphics[width=1\linewidth]{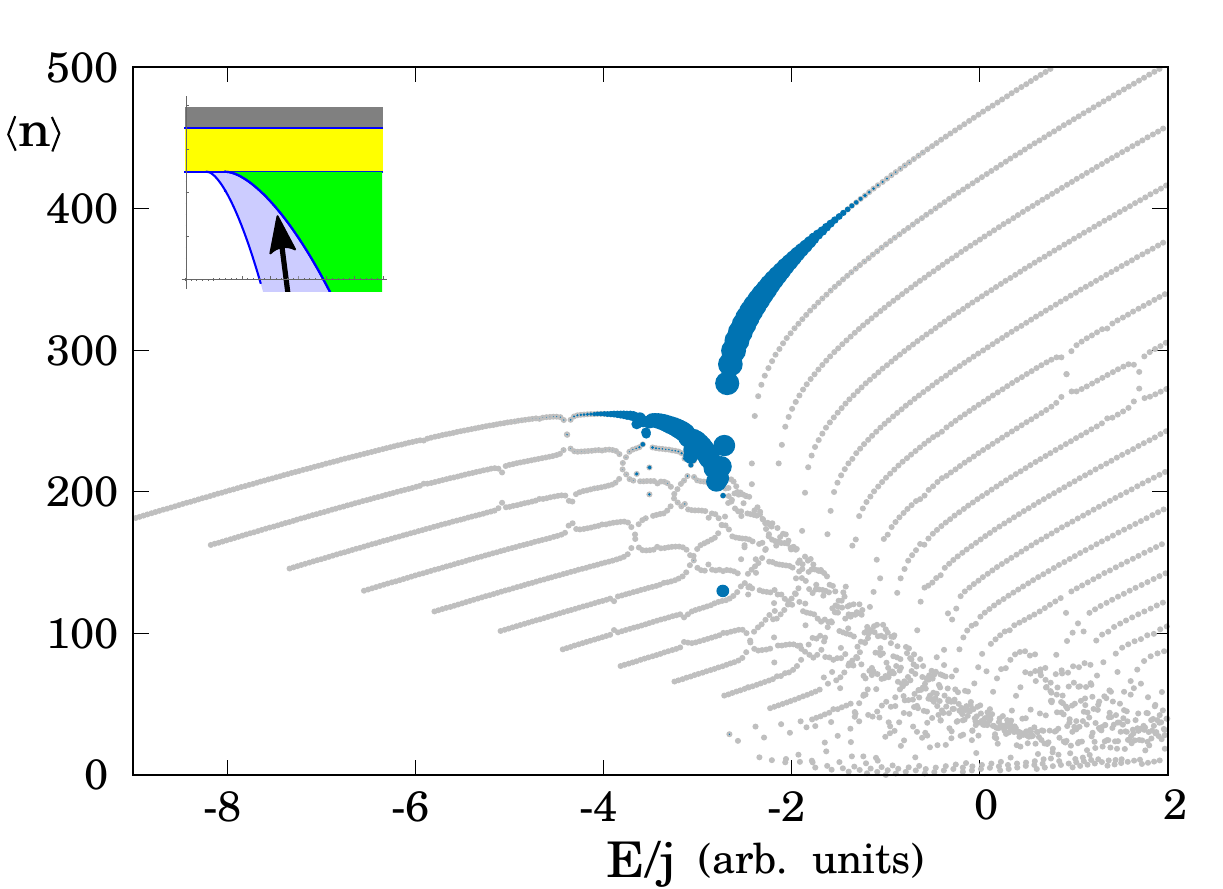}
\caption{ (Color online) The Peres lattice with the average number of photons $\ave{n}$ for the $\lambda \doteq 3.27, \ \delta=0.3$ Hamiltonian (the same parameters as in Fig.\,\ref{F:0p3BQPgs}).
The strength function from Fig.\,\ref{F:0p3BQPgs}(d) is displayed by blue dots, the dot area encodes the respective $|s_l|^2$ value.}
\label{F:Peres0p3}
\end{figure}

All the strength functions in Fig.\,\ref{F:0p3BQPgs} have a common property that their support is only a certain subset of the final Hamiltonian spectrum (see the zero base corresponding to levels which are virtually unpopulated).
This can be interpreted so that the system is in a quasi-regular regime where the overlap with only some selected final states is allowed.
So the modified Heisenberg time $t'_{\rm H}$, restricted only on these states, agrees better with the onset of revivals.

The quasi-regularity of the populated final states is illustrated in Fig.\,\ref{F:Peres0p3} where we present a so-called Peres lattice\,\cite{Per84b} of the final Hamiltonian. 
The Peres lattice depicts the spectrum of eigenstates as a mesh of points in the plane $E_l \times \ave{O}_l$, where $E_l$ is energy and $\ave{O}_l$ an expectation value of a certain observable (here the number of photons) in the $l$th eigenstate.
Orderly arranged points in the lattice indicate regularity of the respective eigenstates whereas disordered points imply chaoticity of eigenstates\,\cite{Str09}.
The states populated in the critical quench from Fig.\,\ref{F:0p3BQPgs}(d) are displayed by the highlighted dots, the size of each dot corresponds to the value $|s_l|^2$ of the strength function.
We observe a localization of the  populated states in the regular domain.
The same is true for the other quenches in Fig.\,\ref{F:0p3BQPgs}. 

\begin{figure}[t]
\includegraphics[width=1\linewidth]{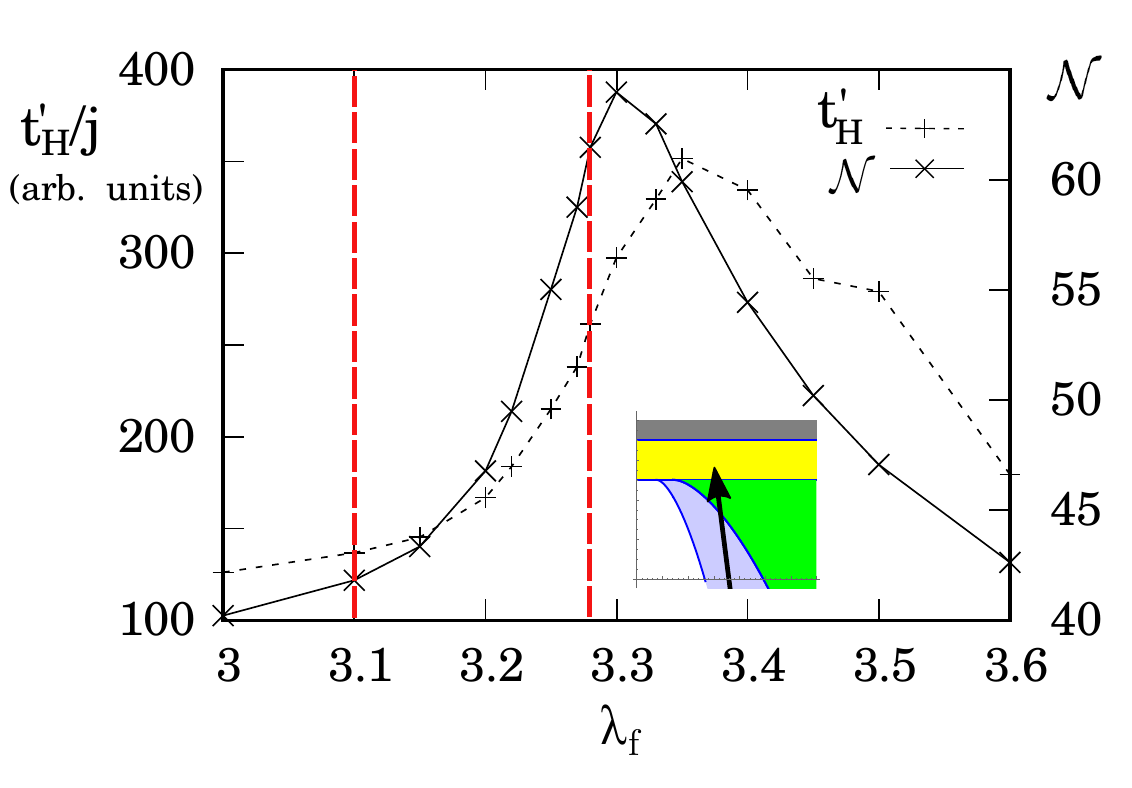}
\caption{ 
(Color online) The modified Heisenberg time $t'_{\rm H}$ and the participation ratio $\mathcal{N}$ for the BQPs in the $\delta=0.3$ model with the $\lambda_{\rm i}=6$ ground state.
The parameters are the same as in Fig.\,\ref{F:0p3BQPgs}.
The critical quenches are marked with vertical lines.
Lines connecting points serve as guides for eyes.
}
\label{F:BQPsmooth0p3}
\end{figure}

In Fig.\,\ref{F:BQPsmooth0p3}, the  modified Heisenberg time $t'_{\rm H}$ and the participation ratio $\mathcal{N}$ are plotted for several values of $\lambda_{\rm f}$.
Both dependences show maxima close to the critical value $\lambda_{\rm f}\doteq 3.27$, in a rough correspondence to the BQPs for $f=1$ critical system (cf. Fig.\,\ref{F:SmoothBack}).
Note that the other critical value $\lambda_{\rm f}=3.1$ induces no effect. 

\begin{figure}[t]
\includegraphics[width=1\linewidth]{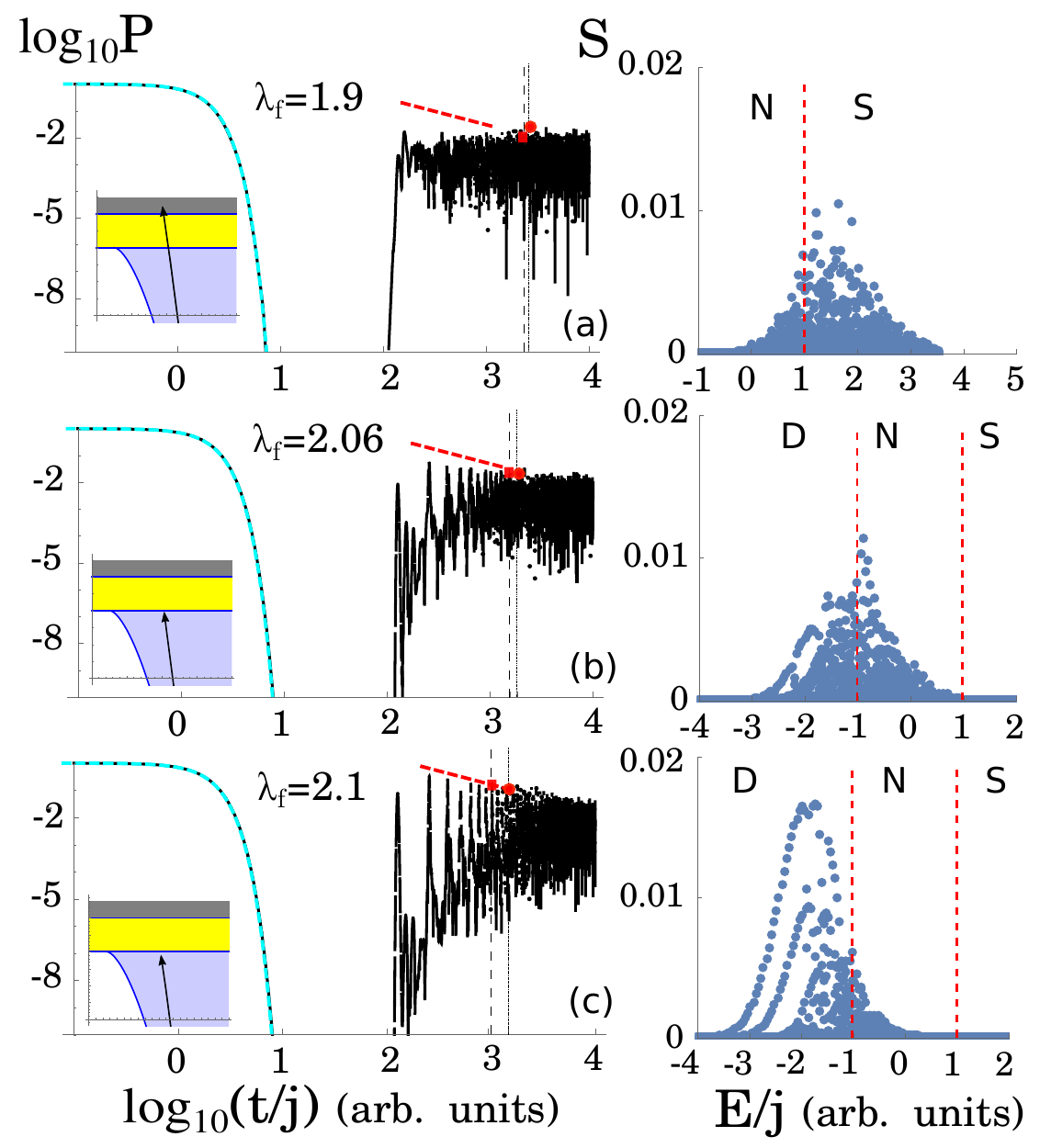}
\caption{ (Color online) The survival probability and the strength function corresponding to the BQPs from $\lambda_{\rm i}=4$ ground state of the $\delta=1$ model.
The settings and symbols are the same as in Fig.\,\ref{F:0p3BQPgs}.
}
\label{F:1p0BQPgs}
\end{figure}

If we increase the parameter $\delta$, the overall degree  of chaoticity involved in the model grows.
Let us move on to probing the quench dynamics in the full Dicke model with $\delta=1$.
In Fig.\,\ref{F:1p0BQPgs}, the survival probability is shown along with the respective strength function for several BQPs from the $\lambda_{\rm i}=4$ ground state.
As  $\lambda_{\rm f}$ we choose three values, with $\lambda_{\rm f}=2.06$ (panel b) corresponding to the critical quench to the $(f,r,)=(2,1)$ ESQPT between the D and N phases at $E=E_{{\rm c}1}$.

As in Fig.\,\ref{F:0p3BQPgs}, the initial decay for the quenches in Fig.\,\ref{F:1p0BQPgs} is Gaussian.
The revivals after the survival collapse can be partially fitted by the $1/t$ envelope in panel\,(c) whereas in panel\,(b) the oscillations are weakened and in panel\,(a) they are not present at all.
This follows from the fact that the corresponding strength functions have much more complex structure than those for $\delta=0.3$.
It is shown in Ref.\,\cite{Ler18} that if the strength function consists of several embedded Gaussian profiles---a clear example is panel (c) of Fig.\,\ref{F:1p0BQPgs}---the interference terms in the survival probability distort the  power-law decay.

\begin{figure}[t]
\includegraphics[width=1\linewidth]{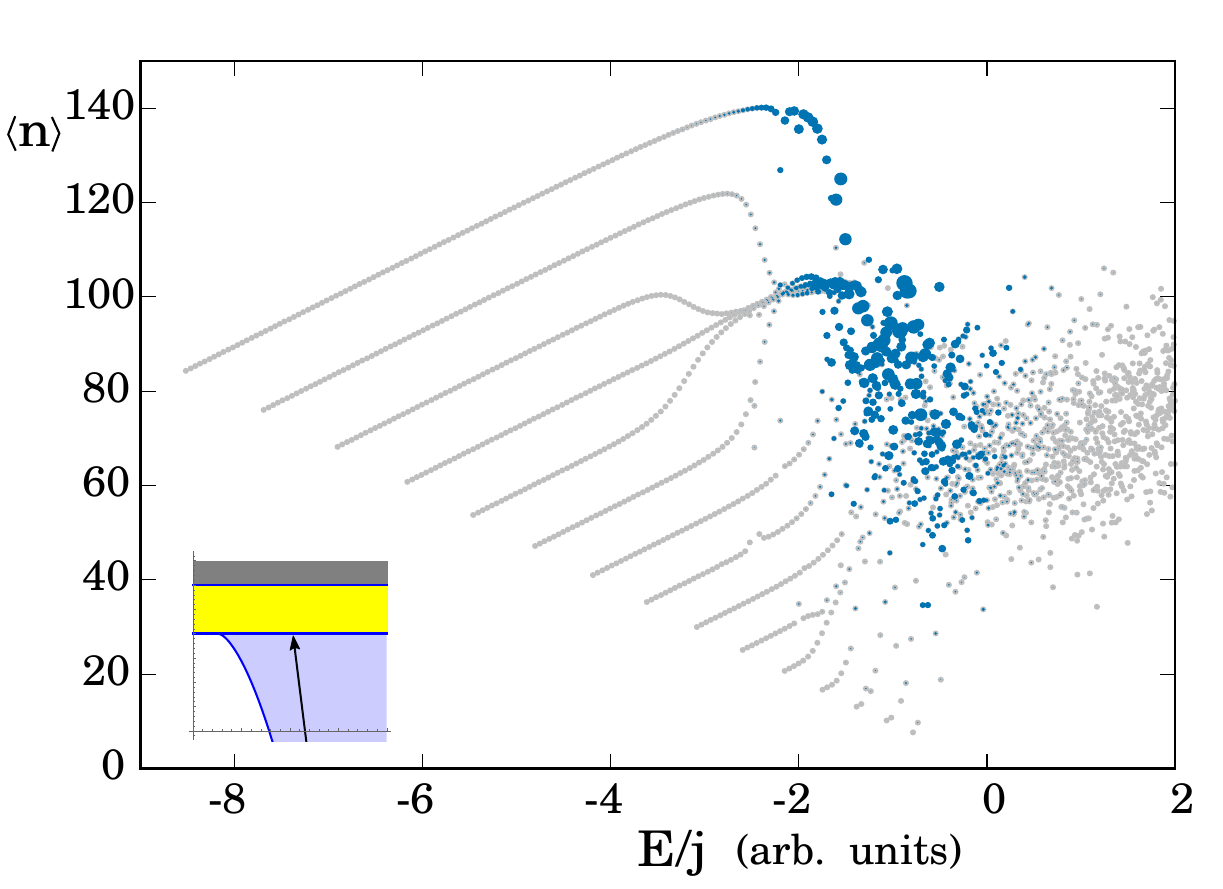}
\caption{ (Color online) The Peres lattice as in Fig.\,\ref{F:Peres0p3} but for the $\lambda=2.06,\ \delta=1$ Hamiltonian.
The strength function from Fig.\,\ref{F:1p0BQPgs}(b) is depicted by blue dots.}
\label{F:Peres1p0}
\end{figure}

The growing complexity of the strength function indicates a higher number of final eigenstates with non-zero $|s_l|^2$.
In Fig.\,\ref{F:Peres1p0} we show the Peres lattice of the final spectrum for the critical quench $\lambda_{\rm f}=2.06$ with the  strength function  encoded in the size of blue points.
We observe that the initial state is distributed mainly in the chaotic part of the spectrum.
This is a radically different situation than for the critical quench to the same ESQPT type  with $\delta=0.3$, cf. Fig.\,\ref{F:Peres0p3}.
Apparently, quantum chaos plays a dominant role in the presently observed disappearance of the power-law modulated revivals at $t \sim t_{\rm H}$,  smearing possible ESQPT effects.
Anyway, in Fig.\,\ref{F:1p0BQPgs}(b)  the presence of the ESQPT between D and N phases is still captured by a partially bimodal form of the critical strength function.

\begin{figure*}[t]
\includegraphics[width=1\linewidth]{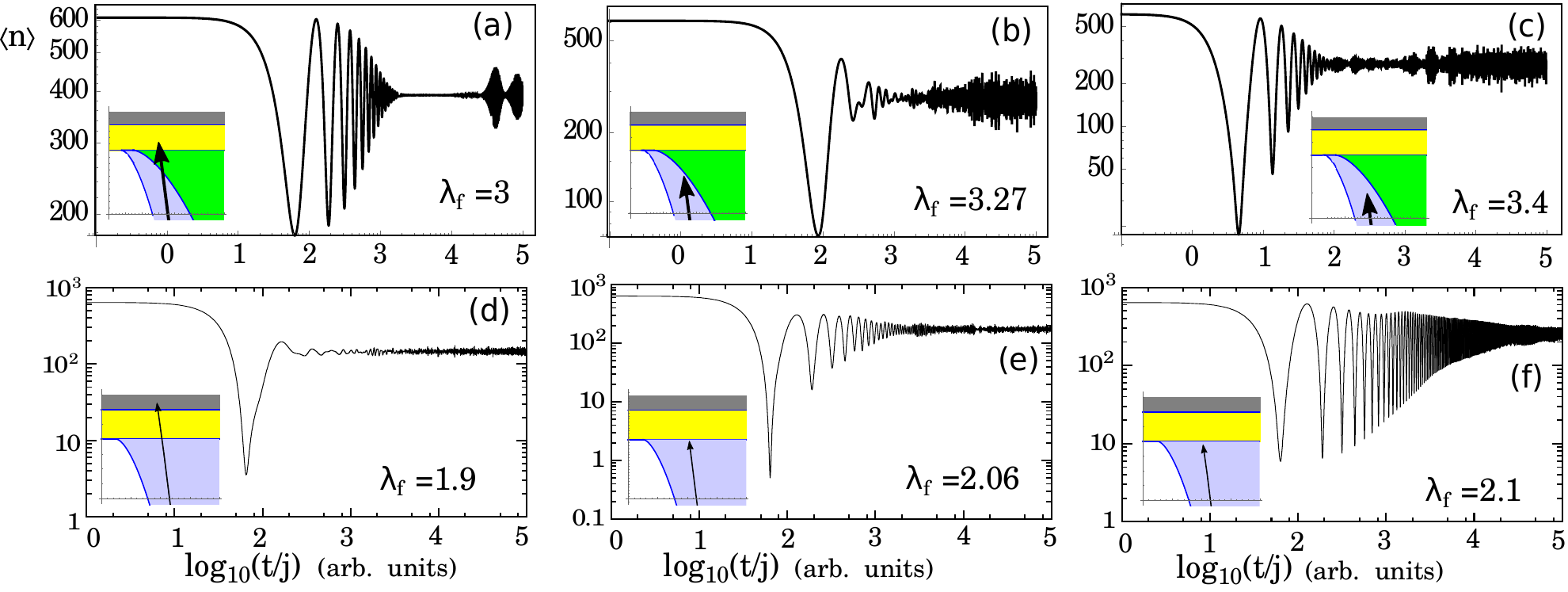}
\caption{ (Color online) The time evolution of the average number of photons in the cavity $\ave{n}$ after the BQPs from the initial  ground state.
Panels (a)--(c) correspond to the $\delta=0.3$ model with  $\lambda_{\rm i}=6$ (the same quenches as in Fig \ref{F:0p3BQPgs}). 
Panels (d)--(f) correspond to the $\delta=1$ model with  $\lambda_{\rm i}=4$ (the same quenches as in Fig \ref{F:1p0BQPgs}). 
Other parameters are  as in Fig.\,\ref{F:0p3BQPgs}.
}
\label{F:OBS}
\end{figure*}

We now attempt to identify some ESQPT-induced effects in the evolution of suitably selected physical observables.
In particular, we look at the 
number of bosons $\hat{n}=\hat{b}^\dagger\hat{b}$, whose average $\ave{n}$ is directly related to the photon flux leaving the cavity in the experimental setup described in Ref.\,\cite{Bau11}.
The evolution of this quantity after the quench can be computed as
\begin{equation}
\ave{n} = \sum_l |s_l|^2 n_{ll}+2 \sum_{l>l'}{\rm Re}\left[s_l s_{l'}^* e^{i(E_{{\rm f}l}-E_{{\rm f}l'})t}\right] n_{ll'},
\label{nt}
\end{equation}
where $n_{ll'}=\langle \phi_{{\rm f}l} |\hat{n}|\phi_{{\rm f}l'}\rangle$.

In Fig.\,\ref{F:OBS} we present results for the above-described BQPs in the $\delta=0.3$ and $\delta=1$ versions of the model.
The $\delta=0.3$ case with $\lambda_{\rm i}=6$ is depicted in panels (a)--(c).
The results are qualitatively similar as in the time evolution of the survival probability, cf. Fig.\,\ref{F:0p3BQPgs}.
In non-critical cases, panels (a) and (c) in Fig.\,\ref{F:OBS}, the oscillations appear after the initial decay.
These are further attenuated, so $ \ave{n}$ reaches its saturation value given simply by the the first term in Eq.\,\eqref{nt}.
In panel\,(b), which corresponds to the critical quench, the oscillatory part of the evolution is suppressed and the saturation regime is reached sooner.
In other words, the time evolution of this observable  captures the presence of ESQPTs in the same way as the survival probability.

The time dependence of $\ave{n}$ after the BQPs in the $\delta=1$ Dicke model with $\lambda_{\rm i}=4$ is plotted in panels (d)--(f) of Fig.\,\ref{F:OBS}.
The critical quench is shown in panel (e).
In analogy to the above described behavior of the survival probability for the same quench protocols, the ESQPT effect in $\ave{n}$ is suppressed due to a high degree of chaoticity of the populated eigenstates of the final Hamiltonian.

\section{Summary} 
\label{SUM}
We  employed various types of quantum quench protocols in multiple settings of the extended Dicke model with the aim to test dynamical signatures of ESQPTs.
Although the information in the time signal is often lost, effects of ESQPTs can be observed in the strength function which is an inverse Fourier transform of the survival probability.
Nevertheless, in the protocols involving the ground states of the initial Hamiltonians, the effect is often visible even in the time dependences.
We observed essentially two types of effects: either the stabilization of the initial state, or a speed-up of its decay.

In the context of the present model, the ESQPT-induced stabilization was observed in the class of forward quench protocols with $\Delta\lambda>0$.
It appears because the final Hamiltonian has a stationary point at the place of the initial Hamiltonian's global minimum.
In our model, the stationary point is stable below the critical coupling ($\lambda_c$ or $\bar{\lambda}_c$) and unstable above (hence inducing an ESQPT).
We examined three different cases:
\begin{itemize}
\item Integrable $\delta=0$ model in its critical $M=2j$ subspace. 
The unstable stationary point affecting the quenches with $\lambda_{\rm f}\in (\bar{\lambda}_c,\infty)$ leads to the logarithmic divergence of the level density as in an ESQPT of the type $(f,r)=(1,1)$.
The stabilization effect was seen in Fig.\,\ref{FQPTC}(c).
\item Non-integrable $\delta\in (0,1]$ model. 
The unstable stationary point affecting quenches with $\lambda_{\rm f}\in (\lambda_c,\lambda_0)$ constitutes an ESQPT with $(f,r)=(2,1)$.
The quench dynamics was shown in Fig.\,\ref{F:0p3FQPgs}(a).
\item Non-integrable $\delta\in (0,1)$ model.
The unstable stationary point affecting quenches with $\lambda_{\rm f}\in [\lambda_0,\infty)$ constitutes an ESQPT with $(f,r)=(2,2)$.
In this case we observed even stronger stabilization due to nearly perfect localization of the strength function, see Fig.\,\ref{F:0p3FQPgs}(b).
\end{itemize}

On the contrary, the ESQPT-induced speed-up of the decay of the initial state was observed in some backward quench protocols.
The initial parameter value  $\lambda_{\rm i}$ was chosen above the critical coupling ($\lambda_c$ or $\bar{\lambda}_c$) and the parameter shift $\Delta \lambda<0$ was set such that the strength function was centered at the ESQPT energy.
The speed-up is manifested as a disappearance or considerable suppression of the power-law stage of the quench dynamics  at long time scales.
The effect was clear in the following cases:
\begin{itemize}
\item Integrable $\delta=0$ model in its critical $M=2j$ subspace, see Fig.\,\ref{FBPTC}(b).
\item Non-integrable $\delta\in (0,1)$ model for quenches to the $(f,r)=(2,1)$ ESQPT separating the D and TC phases, see Fig.\,\ref{F:0p3BQPgs}(d).
 
\end{itemize}

 The presence of the power-law decay at long time scales in the non-critical quenches is due to a combination of (a) Gaussian envelope of the strength function and (b) discrete sampling of the strength function with a quadratic variation of the level spacings.
For the above specified critical quenches, this interplay is violated because of different quadratic dependences of level spacings on both sides of the ESQPT, see Fig.\,\ref{novak} that depicts the situation in the $\delta=0$ model.
Note that in both these cases the Heisenberg time (either $t_{\rm H}$ or $t'_{\rm H}$) and the participation ratio ${\cal N}$ locally increase, see Figs.\,\ref{F:SmoothBack}  and \ref{F:BQPsmooth0p3}.

The suppression of the power-law stage of the quench dynamic is not observed for quenches to ESQPTs of the type $(f,r)=(2,2)$.
Moreover, in the  $\delta=1$ model, the speed-up effect disappears even for $(f,r)=(2,1)$ ESQPT.
This is because the support of the strength function lies in the chaotic part of the final spectrum, cf. Figs.\,\ref{F:Peres0p3} and\,\ref{F:Peres1p0}.

We have demonstrated that similar effects as in the survival probability can be detected in observables like the average photon number in the cavity.
As seen in Fig.\,\ref{F:OBS} this quantity shows a disappearance of the medium-time oscillations for critical quenches to the $(f,r)=(2,1)$ ESQPT in $\delta \in (0,1)$ model.
This may suggest a way of experimental verification of ESQPT-related effects within a cold atom realization of the Dicke-like systems.

  \section{Acknowledgement}
  We acknowledge funding of the Charles University under project UNCE/SCI/013.


\end{document}